\documentclass[twocolumn,10pt]{IEEEtran}
\usepackage{ifpdf, flushend,subfigure}
\ifCLASSINFOpdf
  \usepackage[pdftex]{graphicx}
  \graphicspath{{../pdf/}{../jpeg/}}
  \DeclareGraphicsExtensions{.pdf,.jpeg,.png}
\else
  \usepackage[dvips]{graphicx}
  \graphicspath{{../eps/}}
  \DeclareGraphicsExtensions{.eps}
\fi
\usepackage[hyphens]{url}
\usepackage[dvipsnames]{xcolor}
\usepackage{epstopdf}
\usepackage{multirow}
\usepackage{float}
\usepackage[cmex10]{amsmath}
\usepackage {amssymb}
\usepackage{graphicx}
\usepackage{array}
\usepackage{mdwmath}
\usepackage{mdwtab}
\usepackage{eqparbox}
\usepackage{nomencl}
\usepackage{cite}
\usepackage{algorithm}
\usepackage{algorithmic}
\usepackage{makeidx}
\usepackage{ifthen}
\usepackage{caption}
\usepackage{xurl}
\usepackage{hyperref}
\hypersetup{breaklinks=true}
\usepackage{pdflscape}
\makenomenclature

\renewcommand{\nomgroup}[1]{%
\ifthenelse{\equal{#1}{I}}{\item[\textbf{Indices}]}{%
\ifthenelse{\equal{#1}{A}}{\item[\textbf{Abbreviations}]}{%
\ifthenelse{\equal{#1}{V}}{\item[\textbf{Variables}]}{%
\ifthenelse{\equal{#1}{P}}{\item[\textbf{Parameters and Constants}]}{%
}
}
}
}
}

\newcommand{\argmin}{\operatornamewithlimits{argmin}}

\newcommand{\beq}{\begin{equation}}
\newcommand{\eeq}{\end{equation}}
\newcommand{\beqn}{\begin{eqnarray}}
\newcommand{\eeqn}{\end{eqnarray}}
\newcommand{\beqno}{\begin{eqnarray*}}
\newcommand{\eeqno}{\end{eqnarray*}}
\newcommand{\bma}{\begin{displaymath}}
\newcommand{\ema}{\end{displaymath}}
\newcommand{\bnu}{\begin{enumerate}}
\newcommand{\enu}{\end{enumerate}}
\newcommand{\bce}{\begin{center}}
\newcommand{\ece}{\end{center}}
\newcommand{\btb}{\begin{tabular}}
\newcommand{\etb}{\end{tabular}}

\begin{document}

\title{A Bilevel Programming Framework for Joint Edge  Resource Management and Pricing}


\author{\IEEEauthorblockN{Tarannum Nisha,~\IEEEmembership{Student Member,~IEEE}, Duong Tung Nguyen,~\IEEEmembership{Member,~IEEE}, \\and Vijay K.  Bhargava,~\IEEEmembership{Life~Fellow,~IEEE}}  
\thanks{Tarannum Nisha and Vijay K. Bhargava are with the Department
of Electrical and Computer Engineering, University of British Columbia,
Vancouver, Canada.
Email: \textit{tarannum.abu@gmail.com}, \textit{vijayb@ece.ubc.ca. }
Duong Tung Nguyen is with the School of Electrical, Computer, and Energy Engineering, Arizona State University, Tempe, United States.
Email:  \textit{duongnt@asu.edu.} 
This research was supported, in part, by the Natural Sciences and Engineering Research Council of Canada.
(\textit{Corresponding author: Duong Tung Nguyen}).
 }
 }



\maketitle
\begin{abstract}
The emerging edge computing paradigm promises to provide low-latency and ubiquitous computation to numerous mobile and Internet of Things (IoT) devices at the network edge. How to efficiently allocate  geographically distributed heterogeneous edge resources to a variety of services 
is a challenging task. 
While this problem has been studied extensively in recent years,  most of the previous work has 
largely ignored 
the preferences of the services when making edge 
resource allocation decisions. 
To this end, this paper introduces a novel bilevel optimization model, 
 which  explicitly takes the service preferences into consideration, 
to study the interaction between an EC platform and multiple services. 
The  platform manages a set of edge nodes (ENs) and acts as the leader while the services are the followers.  Given the service placement and  resource pricing decisions of the leader, each service decides how to optimally divide its workload to different ENs.
The proposed framework not only maximizes the profit of the platform but also minimizes the cost of every service. When there is a single EN, 
we derive a simple analytic solution for the underlying problem. For the general case with multiple ENs and multiple services, we present a  Karush–Kuhn–Tucker-based solution and a duality-based solution, combining with a series of linearizations, to solve the  bilevel problem. 
Extensive numerical results are shown to illustrate the efficacy of the proposed model.

\end{abstract}


\printnomenclature

\section{Introduction}


Edge computing (EC) has emerged as a vital technology that 
works in tandem with the cloud to  mitigate network traffic, improve user experience, and enable various
IoT applications.
By distributing computational and storage resources 
to the proximity of end-users and data sources, the new computing paradigm 
offers remarkable 
capabilities, such as local data processing and analytics, resource pooling and sharing,  real-time computing and learning, enhanced security and reliability, distributed caching, and localization. 
Additionally, EC  
is 
the key to satisfying the stringent requirements of exciting new systems and low-latency 
applications such as virtual/augmented reality (VR/AR),  embedded artificial intelligence, autonomous driving, manufacturing automation,  and tactile Internet. In  future networks, edge resources form an intermediary layer between multitude of diverse but constrained end-devices and the large cloud data centers (DCs)  \cite{wshi16}.

Despite the rapid growth witnessed in  EC technology and tremendous potential it holds for upcoming years, it is still in its infancy stage and many challenges remain to be addressed. 
One of the most important challenges is the problem of 
multi-tenancy of shared and heterogeneous edge resources, which is also the main focus of this paper. 
In particular,  we study the interaction between an EC platform 
and multiple services (e.g., AR/VR applications, Google Maps). The platform (e.g., a telco,  a cloud provider, 
a third-party \cite{equinix}) manages a set of ENs 
and can monetize the edge resources by selling them to the services. 
By placing and running the services at the ENs, the service providers (SPs) can drastically enhance the quality of experience for their users since the user requests can be served directly at the network edge.

Our work aims to address
  two fundamental questions:  \textit{(1) how can the platform   set the edge resource prices optimally}, and \textit{(2) how much resources should a service  purchase} 
\textit{from each EN.} These  questions are challenging due to the interdependence between the decisions of the platform and the services. Specifically, the resource procurement decisions of the services depend on the  resource prices set by the platform.  On the other hand, the pricing decisions of the platform depends on the resource demands of the services. 

Also, because of the heterogeneity of the ENs, the services may have diverse preferences towards them. Consequently, the valuations of different ENs to a service can be different.  In general, a service prefers low-priced edge resources as well as  ENs with powerful hardware and  geographically  close  to  it. 
To minimize the network delay between its users and computing nodes, a service tends to procure  resources from its closest ENs. Hence, some ENs can be over-demanded (e.g., ENs in or  near high-demand areas) while some other nodes are under-demanded, which leads to low resource utilization. Intuitively, the platform can  reduce the resource prices of underutilized ENs to encourage load shifting  from the overloaded ENs.

To this end, we formulate a joint edge resource management and pricing  problem between the platform and the services, and propose to cast it as a bilevel optimization model \cite{asin18} (i.e., a Stackelberg game). The proposed model can not only assist the platform to determine the optimal edge resource prices to maximize its profit, but also help each service  find an optimal resource procurement and workload allocation solution to minimize its cost while improving the user experience. 
In the formulated bilevel problem, the platform is the leader and each service is a follower. The leader decides the optimal resource prices to assign to different ENs, while anticipating the reaction of the followers. 
Given the edge resource prices computed by the leader, each service solves a follower problem to 
identify the optimal amount of resource to buy from each EN, considering its delay and budget constraints.

To the best of our knowledge, this is the first bilevel programming formulation for the joint edge service placement, resource procurement, and pricing problem. 
Note that while Stackelberg games have been used extensively to study various problems in EC, most of the existing models contain a simplified follower problem that normally has a special closed-form solution to facilitate the backward induction method. For our problem, we followed a similar procedure   to obtain an analytic solution for the case with  a single EN in the system. However, for the general case with multiple ENs, the follower problem becomes sophisticated. Also, our formulation contains integer service placement variables. Hence, 
backward induction 
cannot be applied to solve our bilevel optimization problem. Our formulation makes it easier and more flexible for the services to express their objective functions and  constraints.

Bilevel optimization problems are  generally extremely hard to solve \cite{asin18}. In this paper, we present two solutions to compute an exact optimal solution to the formulated  mixed integer non-linear bilevel program (MINBP) in the general case. The first solution relies on  the Karush–Kuhn–Tucker (KKT) conditions to convert the  bilevel problem into an equivalent mathematical program with equilibrium constraints (MPEC) \cite{sgab12}, which is a mixed integer nonlinear program (MINLP). By employing the strong duality theorem and some  linearization techniques, we  transform this MINLP into a mixed integer linear program (MILP) that can be solved efficiently using off-the-shelf MILP solvers such as Gurobi\footnote{https://www.gurobi.com/} 
and Cplex\footnote{https://www.ibm.com/analytics/cplex-optimizer}. 

Although the first solution can solve the bilevel program optimally, the resulting MILP has a large number of constraints and auxiliary integer variables due to the complimentary slackness constraints from the KKT conditions.
Therefore, we propose an alternative solution that uses linear programming (LP) duality and a series of linearizations  to convert the original bilevel problem into an equivalent MILP with  significantly less number of constraints and integer variables 
compared to the one obtained by the KKT-based approach.
Our main contributions can be summarized as follows:



\begin{itemize}
    \item \textit{Modeling:} We propose a novel bilevel optimization framework for joint edge resource management and pricing, where the platform optimizes the  resource pricing, EN activation, and service placement decisions in the upper level while each service optimizes its workload allocation decisions in the lower level. The service preferences are explicitly captured in the proposed framework. 
    
    \item \textit{Solution approach:} The formulated problem is a challeging MINBP. We first present an analytic solution for the special case with a single EN. When there are multiple ENs in the system, we develop two efficient approaches based on the KKT conditions and LP duality, respectively, to optimally solve the bilevel problem.
    
    \item \textit{Simulation:} Extensive numerical results are shown to illustrate the effectiveness of the proposed scheme, 
   which provides a 
win-win solution for both the EC platform and the services. In
particular, it can help increase the  profit for the
platform, decrease the costs for the services, and  improve the edge resource utilization.
\end{itemize}


The rest of the paper is organized as follows. The system model is described in Section \ref{model}. Section \ref{for} introduces the problem formulation. The solution approaches and simulation results are presented in Section \ref{sol} and Section \ref{sim}, respectively.  Section \ref{rel} discusses related work followed by the conclusions and future work  in Section \ref{conc}.

\section{System Model}
\label{model}

We consider a system that consists of an EC platform, also known as an edge infrastructure provider, and a set $\mathcal{K}$  of $K$ 
services. 
The platform manages a set $\mathcal{N}$ of $N$ ENs 
to provide computational resources to the services. 
The services can be proactively installed onto selected ENs to  reduce the communication latency and improve  service quality. 
In practice, various sources (e.g., under-utilized  DCs in schools/malls/enterprises, idle PCs in research labs,  edge servers at base stations, telecom central offices) can serve as ENs \cite{wshi16}.
In addition to its own ENs, the platform may also control  ENs owned by other entities (e.g., telcos, malls, universities). The EN owners can offer their idle edge resources to the platform in exchange for a certain compensation. 
The service requests from end-devices normally arrive at a point of  aggregation (e.g., switches, routers, base stations, WiFi access points), then  will be forwarded to an EN or the  cloud for processing.  

\begin{figure}[hbt!]
	\centering
		\includegraphics[width=0.45\textwidth,height=0.22\textheight]{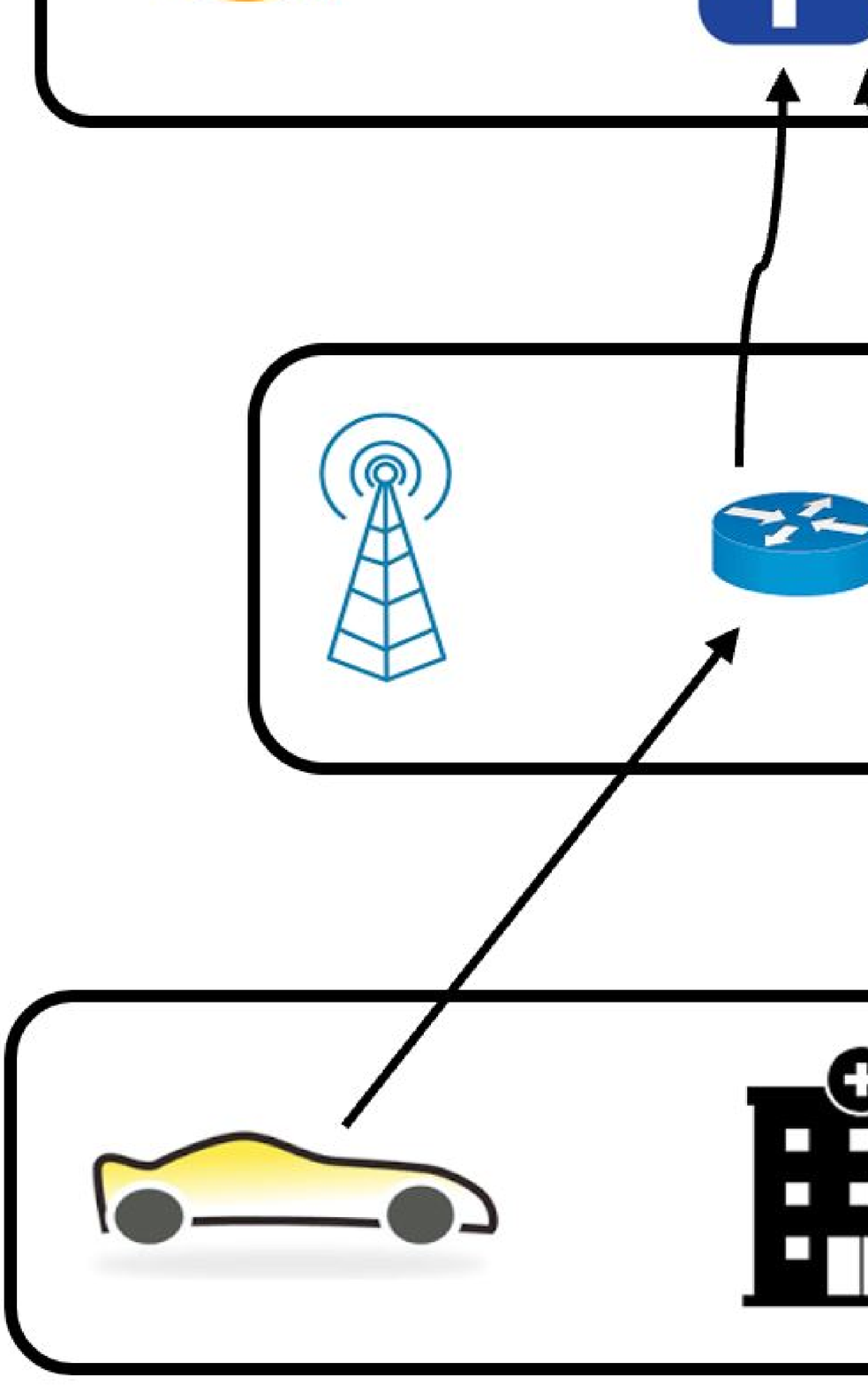}
			\caption{System model}
	\label{fig:model}
\end{figure}

Throughout this paper, the points of aggregation are referred to as access points (AP). 
We assume 
there is a set $\mathcal{M}$ of $M$ APs in the system. Each service serves users located in different areas, each of which is represented by an AP. Note that an EN can be co-located with an AP. For instance, edge servers can be placed at a base station. In enterprise data centers, edge servers can be deployed near routers/switches.  
Similar to the previous literature \cite{duongtcc,duongton,duongiot,vfar19,lwan18,mjia17,syan19,ayou19,kpout19}, we  study the service placement and request scheduling problem from the APs to the ENs and the  cloud only.
Fig.~\ref{fig:model} depicts the system model.

Let $i, j$, and $k$  be the AP index, EN index, and service index,  respectively.
The network delay between AP $i$ and EN $j$ is $d_{i,j}$, and the delay between AP $i$ and the cloud is $d_{i,0}$.
The goal of each service is to minimize not only the resource procurement cost but also the  network delay for its users. 
Define $R_i^k$ as the resource demand (i.e., workload) of service $k$ at AP $i$. Given the  resource prices, locations, and specifications of the ENs, each service decides how to optimally divide its workload to the active ENs and the  cloud for processing. 
In  Fig.~\ref{fig:model}, each active EN is represented by a green dot while a red dot indicates an inactive EN.


Each service $k$ has a budget $B^k$ for  resource procurement. 
The amount of workload of service $k$ at AP $i$ assigned to EN $j$
is denoted by $x_{i,j}^k$.
Also, $x_{i,0}^k$ is the amount of workload of service $k$ at AP $i$  routed to the cloud. 
Define $x_0^k = (x_{1,0}^k, x_{2,0}^k, \ldots, x_{M,0}^k)$, $x_i^k = (x_{i,1}^k, x_{i,2}^k, \ldots, x_{i,N}^k)$, and $x^k = (x_1^k, x_2^k, \ldots, x_M^k)$. 
Clearly, to enhance the user experience, a service 
prefers to have its requests  processed by ENs closer to its users rather than the remote cloud.  Let $y_0^k$ and $y_j^k$
represent the amounts of computing resources that service $k$
 purchases from the cloud and EN $j$, respectively. Also, $y^k = (y_1^k, y_2^k, \ldots, y_N^k)$. Define $D^{k, \sf m}$ as the maximum delay threshold of service $k$. The average delay of service $k$ at AP $i$ is $d_{i}^{k, \sf a}$. Denote by $w^k$  the delay penalty parameter for service $k$. Let $s^k$ be the  size of service $k$. 
The placement cost of service $k$ at EN $j$ is $\phi_j^k$, which includes the  downloading, installation, and storage costs. 
The binary variable $t_j^k$ indicates if service $k$ is installed at EN $j$ or not. Define $t^k = (t_1^k, t_2^k, \ldots, t_N^k)$ and $t = (t^1, t^2, \ldots, t^K)$.

For each EN $j$, its storage capacity and computing capacity are denoted by  $S_j$ and $C_j$, respectively.  Since the services may have different preferences  towards the ENs, some ENs can be over-demanded while others are under-demanded. Hence, a natural solution is to efficiently price the edge resources to balance supply and demand.  The unit price of computing resource of EN $j$ is denoted by  $p_j$. Specifically, $p_j$ is the amount of money  that each service needs to pay to the platform for renting one computing unit at EN $j$ during the whole scheduling period.
Define $p = (p_1, p_2, \ldots, p_N)$.
Moreover, due to the limited storage resource, each EN can host only a subset of services. 
The operating cost of 
an active EN $j$ includes a fixed cost $c_j$ and a variable cost $q_j$ depending on its computing resource utilization. Let $z_j$ be a binary  variable that  equals one if EN $j$ is active and zero otherwise. Define $z = (z_1, z_2, \ldots, z_N)$. 
The platform needs to jointly decide which ENs to activate, which service to place at which node, and the resource prices of  individual ENs to maximize its revenue while minimizing the total operation cost. 
The main notations are summarized in Table \ref{notation}.

\begin{table}[ht!] 
\centering
\caption{NOTATIONS}
\begin{tabular}{|l|l|}
\hline
Notation   & Meaning\\	
\hline	
$i$, $M$, $\mathcal{M}$ & Index, number, and set of APs\\
\hline
$j$, $N$, $\mathcal{N}$  & Index, number, and set of ENs \\
\hline	
$k$, $K$, $\mathcal{K}$ &  Index, number, and set of services\\
\hline
$S_j$ & Storage capacity of EN $j$\\
\hline
$C_j$ & Computing resource capacity of EN $j$\\
\hline
$c_j$, $q_j$ & Fixed cost and variable  cost of  EN $j$ \\
\hline
$d_{i,j}$ & Network delay between AP $i$ and EN $j$\\
\hline
$d_{i,0}$ & Network delay between AP $i$ and the cloud\\
\hline
$D^{\sf t}$ & Network  access  delay  threshold\\
\hline
$D^{k, \sf m}$ &   Delay threshold of service $k$ \\
 \hline
$d_{i}^{k, \sf a}$ &   Average delay  of service $k$ in area $i$ \\
 \hline
$B^k$ & Budget of service $k$  \\
\hline
$s^k$ & Storage resource requirement of service $k$\\
\hline
$w^k$ & Delay penalty parameter for service $k$ \\
\hline
$R_i^k$ & Resource demand of service $k$ at AP $i$\\
\hline
$\phi_j^k $ & Placement cost of service $k$ at EN $j$\\
\hline
$t_j^k$ &  Binary variable, 1 if service $k$ is placed at EN $j$  \\
\hline
$z_j$ & Binary variable, 1 if EN $j$ is active\\
\hline
$p_j$ & Unit price of computing resource at EN $j$ \\
\hline
$x_{i,j}^k$ & Workload of service $k$ at AP $i$ assigned to EN $j$\\
\hline
$x_{i,0}^k$ & Workload of service $k$ at AP $i$ assigned to the cloud \\
\hline
$y_j^k$ & Amount of  resource  of EN $j$ allocated to service $k$\\
\hline
$y_0^k$ & Amount of  resource  of cloud allocated to service $k$\\
\hline

\end{tabular} \label{notation}
\end{table}

Our work focuses on the interaction between the platform and multiple latency-sensitive services. The platform needs to properly price resources at different ENs to maximize its profit and ensure load balancing, while considering diverse service preferences.
The edge resource prices  are interdependent because whether  a service chooses to offload its tasks to an EN or not depends not only on the price at that EN but also on the prices at other ENs. Besides resource pricing, the platform is also responsible for downloading and installing the services onto different ENs. The placement decision is subject to  the storage capacity constraints of the ENs.

By anticipating the reaction of the services, the platform optimizes the resource prices and service placement. 
Given the pricing and placement decisions announced by the platform, each service  responds  by computing its favorite 
edge resource bundle (i.e., the optimal amount of resource to purchase from each EN).
Since the platform acts first and the services make their decisions based on the platform's decisions, the 
process is
sequential. Thus, we model the interaction between the platform and the services as a bilevel program, where the platform and services are the leader and followers, respectively.


\section{Problem Formulation}
\label{for}


\begin{figure}[hbt!]
	\centering
		\includegraphics[width=0.45\textwidth,height=0.14\textheight]{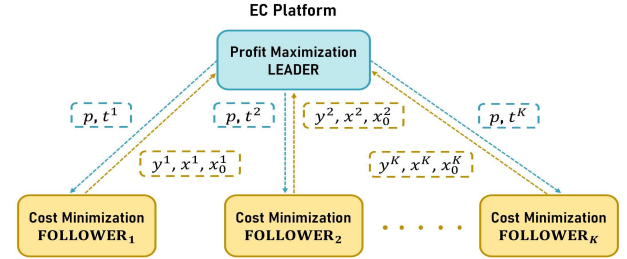}
			\caption{Interaction between the platform and services}
	\label{fig:interaction}
\end{figure} 
In this section, we formulate the interaction between the platform (i.e., the leader) and the services (i.e., the followers) as a bilevel program which consists of an upper-level optimization problem and $K$ lower-level problems, each for one service. The platform solves the upper-level  problem to 
maximize its profit, and then announces the  resource prices 
and service placement decisions to the services. After receiving the information from the platform, each service solves a lower-level  problem to minimize its cost under the delay and budget constraints, 
and then send the optimal resource procurement and workload allocation solution  back to the platform. 

 Fig.~\ref{fig:interaction} summarizes the interaction between the platform and services. In bilevel programming,  the upper-level problem is commonly referred to as the leader problem while the lower-level ones are the follower problems. The optimal solutions of the followers and the leader are interdependent. In particular, the decisions of the followers serve as input to the profit maximization problem of the leader. The output of the leader problem also directly affects the followers' decisions. The follower problems are indeed  constraints to the leader problem. 
In the following, we will describe the follower problem for each service, the leader problem for the platform, as well as the bilevel optimization model.




\subsection{The Follower Problem}

Given the resource prices and service placement decisions announced  by the platform, each service aims to minimize not only the resource procurement  cost but also the total network delay by judiciously distributing its workload to the cloud and the ENs that have installed the service. The  cost of service $k$ for purchasing cloud resource is $p_0 y_0^k$, where $p_0$ is the unit resource price    at the cloud. 
The total cost of service $k$ for purchasing edge resources is $\sum_j p_j y_j^k$. Thus, the total resource procurement cost for service $k$ is $p_0 y_0^k + \sum_j p_j y_j^k$. The delay cost between AP $i$ and EN $j$ is proportional to the amount of workload allocation from AP $i$ to EN $j$ and the network delay between them. Hence, the delay cost of service $k$ can be expressed as $w^k (\sum_i x_{i,0}^k d_{i,0} + \sum_{i,j} x_{i,j}^k d_{i,j}) $. 
The goal of  service $k$ is to minimize the following objective function, which is the sum of  its resource cost and delay cost:
\beqn
 p_0 y_0^k + \sum_j p_j y_j^k +  w^k \Big(\sum_i x_{i,0}^k d_{i,0} + \sum_{i,j} x_{i,j}^k d_{i,j} \Big),
\eeqn
where the delay penalty parameter $w^k$  can be adjusted by the service  to control the tradeoff between the resource procurement cost and the delay cost. A higher value of $w^k$ implies that the service is more delay-sensitive and willing to pay more to buy edge resources to reduce its overall delay. Intuitively, when $w^k$ is high, the service  tends to buy more edge resources to lower the delay penalty cost. Thus, depending on its attitude towards the delay, which directly impacts the user experience, the service can choose a suitable value of $w^k$. Note that the actual payment of each service is the resource procurement cost only. The delay penalty cost is a virtual cost, which is used to express the delay-sensitive level of the service.

 The constraints of the follower problem are described in the following.
First, the total workload of service $k$ allocated to  EN $j$ cannot exceed the amount of computing resource purchased from the EN, i.e., we have: 
\beqn
\label{eq5}
 \sum_i x_{i,j}^k \leq  y_j^k, ~~ \forall j, k.
\eeqn
Similarly, for the resource purchased from the cloud, we have:
\beqn
\label{eq6}
  \sum_i x_{i,0}^k \leq  y_0^k, ~~ \forall k.
\eeqn
The resource demand from AP $i$ must be served by either the cloud or some EN, which implies:
\beqn
\label{eq7}
x_{i,0}^k + \sum_j x_{i,j}^k = R_i^k,~~ \forall i, k. 
\eeqn
While the capacity of the cloud is virtually unlimited, the resource of each EN is limited. Hence, the amount of resource purchased from an EN cannot exceed the capacity of that node. Additionally, service $k$ buys resources from EN $j$ only if the service is placed on EN $j$ (i.e., $t_j^k = 1$). Therefore: 
\beqn
\label{eq51}
  y_j^k \leq C_j t_j^k, \quad \forall j,k.
\eeqn
Since the total resource procurement cost of a service is limited by its budget, we have:
\beqn
p_0 y_0^k  + \sum_j p_j y_j^k  \leq B^k, ~~ \forall k.
\eeqn
The average delay of service $k$ in area $i$ can be expressed as: 
\beqn
d_{i}^{k, \sf a} = \frac{x_{i,0}^k d_{i,0} +  \sum_j x_{i,j}^k d_{i,j}}{ R_i^k},~~\forall i,k.
\eeqn
For a delay-sensitive service, it may require that the average delay in every area should not exceed a certain delay threshold, which imposes $d_{i}^{k, \sf a} \leq D^{k, \sf m},~\forall i,k$.

Furthermore, each service may have certain 
 hardware and software requirements for the ENs that can host the service.
For example, some service can only be deployed on
ENs that support TensorFlow and Ubuntu. Additionally,  if a service  is delay-sensitive, its requests from any area should be
handled by ENs that are not too far from that area.
Thus, we use a binary indicator
$a_{i,j}^k$ to indicate whether EN $j$  
is eligible to serve the demand of service $k$ at AP $i$ or not. The parameters $a_{i,j}^k$ can be set either manually or automatically.
Clearly, we have: 
\beqn
\label{eq72}
 x_{i,j}^k \leq a_{i,j}^k R_i^k,~~ \forall i,j,k.
\eeqn

Overall, the follower problem for service $k$ can be written as follows:
\beqn
\label{objf}
\underset{x^k,y^k}{\text{min}} ~p_0 y_0^k +  \sum_j p_j  y_j^k +  w^k \Big(\sum_i x_{i,0}^k d_{i,0}  + \sum_{i,j} x_{i,j}^k d_{i,j}\Big)
\eeqn
~subject to 
\beqn
\label{eef1}
   x_{i,0}^k + \sum_j x_{i,j}^k = R_i^k, ~~ \forall i \quad (\xi_i^k)\\
\label{eef3}
    \sum_i x_{i,0}^k \leq y_0^k, \quad (\mu_1^k)\\
\label{eef4}
     \sum_i x_{i,j}^k \leq y_j^k, ~~ \forall j \quad (\lambda_j^k)\\
\label{eef41}
     y_{j}^k \leq C_j t_j^k, ~~ \forall j \quad (\Gamma_j^k)\\
\label{eef5}
x_{i,j}^k \leq a_{i,j}^k R_i^k,~~ \forall i,j  \quad (\eta_{i,j}^k) \\
\label{eef6}
\sum_j p_j y_j^k + p_0 y_0^k \leq B^k \quad (\mu_2^k)\\
\label{eef2}
    x_{i,0}^k d_{i,0} + \sum_j x_{i,j}^k d_{i,j} = d_{i}^{k, \sf a} R_i^k,  ~~\forall i \quad (\sigma_i^k)\\
\label{eef21}
    d_{i}^{k, \sf a} \leq D^{k, \sf m},  ~~\forall i \quad (\tau_i^k)\\
\label{eef7}
    x_{i,0}^k \geq 0, ~~\forall i \quad (\zeta_{i,0}^k )\\
\label{eef71}
    x_{i,j}^k \geq 0,~~ \forall i,j, \quad (\epsilon_{i,j}^k)
\eeqn
where the notations in the parentheses associated with the constraints are the  Lagrange multipliers 
of the corresponding constraints. It is easy to see from the follower problem (\ref{objf})-(\ref{eef71}) that a service  buys resource from an EN only if the gain from delay reduction outweighs  the cost increment due to the price difference  between the cloud resource and edge resource. Note that we have $K$ follower problems, one for each service. In addition, although  $p$ and $t$ are  variables in the  leader problem, they are parameters in the follower problems.

\subsection{The Leader Problem}

The objective of the platform is to maximize its profit which is equal to revenue minus  cost. 
The revenue of the platform obtained from selling computing resources  is $\sum_j p_j \sum_k y_j^k$, where 
$\sum_k y_j^k$ is the total amount of computing resource from EN $j$ allocated to the services. The total cost of the platform includes the operating cost of the ENs and the service placement cost. 
The operating cost of an EN depends on the electricity price and power consumption of the node. For simplicity, as commonly assumed in the literature \cite{linear, linear1},  the operating cost of a node is approximated by a linear function. 
When an EN is active, its operating cost is the sum of a fixed cost and a variable cost which  depends on its computing resource utilization. 
Thus, the operation cost of EN $j$ can be expressed as: 
\beqn
\label{ecost}
 Cost_j^{\sf e} = c_j z_j + q_j  \frac{\sum_{i,k} x_{i,j}^k}{C_j},~~ \forall j.
\eeqn
The second term is actually  $q_j  z_j \frac{\sum_{i,k} x_{i,j}^k}{C_j}$. However,  we later enforce that if EN $j$ is not active  (i.e., $z_j = 0$), then $y_j^k = 0, \forall k$. Hence, from (\ref{eq5}), we can ignore $z_j$ in the second term. 
Note that if EN $j$ is owned by a third party, we can simply set $q_j = 0$ in (\ref{ecost}), and interpret $c_j$ as the price of the  EN offered by the third party and $z_j$ as a binary indicator which equals one if the platform buys EN $j$ and zero otherwise. 

Besides the electricity cost, in this work, we consider the setting where the platform is also responsible for the service placement cost.  The placement cost $\phi_j^k$ captures the downloading, installation, and storage costs of service $k$ at EN $j$.
Since a service can only operate on an active EN, the  cost of running service $k$ on EN $j$ is $\phi_j^k t_j^k,~ \forall j,k.$

Overall, the profit of the platform is:
\beqn
\label{profiteq}
\mathcal{P} =  \sum_{j,k} p_j  y_j^k - \sum_j \Bigg[ c_j z_j +  q_j  \frac{\sum_{i,k} x_{i,j}^k}{C_j} + \sum_k \phi_j^k t_j^k \Bigg].
\eeqn



Next, we describe the leader problem's constraints. 
The EN activation and service placement decisions are binary. Thus:
\beqn
\label{leaderc0}
t_j^k \in \{0,~1\},~\forall j,~k;~~z_j \in \{0,~1\}, ~\forall j.
\eeqn
Since a service can only be installed on active ENs, we have:
\beqn
\label{leaderc1}
    t_j^k \leq z_j,~~ \forall j,~k.
\eeqn
We can only allocate computing resource from an active EN to the services. Furthermore, the total allocated  computing resource from an EN cannot exceed its computing capacity. Therefore, we have:
\beqn
\label{leaderc2}
    \sum_k y_j^k \leq z_j C_j, ~~ \forall j,
\eeqn
which implies if $z_j$ = 0, then $y_j^k = 0, ~\forall j,k$. Hence, the services cannot receive computing resource from an inactive EN.
Similarly, the total storage resource of an EN allocated to the services is limited by its storage capacity, i.e., we have:
\beqn
\label{leaderc3}
     \sum_k s^k t_j^k \leq z_j S_j, ~~ \forall j,
\eeqn
where $s^k$ is the storage size required for storing   service $k$.

We assume that the unit resource price at each EN belongs to a predefined discrete set, i.e., we have
\beqn
    p_j \in \{p_j^1 , \ldots, p_j^V \}, ~ \forall j,
\eeqn
where  $v \in \{1, \ldots, V\}$ represent different price options $p_j^1 < p_j^2 < \ldots < p_j^V$. 
This is a natural assumption since the price options can express different levels of the price (e.g., very low price, low price, medium price, high price, very high price). Another reason that we use discrete sets to express the prices is due to the linearization procedure described later in the solution approach section. Note that if the price is continuous, we can  discretize the price range into a large number of intervals. Please refer to \textit{Appendix \ref{discreteAppen}} for more details. Let $r_j^v$ be a binary variable which  equals one if the resource price at EN $j$ is  $p_j^v$. Since the price can take only one value, we have:
\beqn
\label{leaderc4}
p_j = \sum_v p_j^v r_j^v,~\forall j; ~ 
\sum_v r_j^v = 1,~\forall j; ~r_j^v\in\{0,1\}, \forall j, v.
\eeqn

We are now ready to present the leader problem, which is indeed a bilevel program as presented below: 
\beqn
\label{objlead}
\underset{p,z,t,x,y}{\text{max}}  \sum_{j,k} p_j  y_j^k -  \Bigg[ \sum_j  \big(c_j z_j +  q_j  \frac{\sum_k y_j^k}{C_j} \big)
+ \sum_{j,k}  \phi_j^k t_j^k  \Bigg]
\eeqn
subject to
\beqn
&&   (\ref{leaderc0}) - (\ref{leaderc4})  \nonumber \\
\label{eel5}
\label{fol}
  &&  x^k, y^k\in \underset{(x^k, y^k)  \in \mathcal{F}^k}{\argmin}  \Bigg\{  \sum_j p_j  y_j^k + p_0 y_0^k + \nonumber \\
&& \quad \quad w^k \Big(\sum_i x_{i,0}^k d_{i,0} + \sum_{i,j} x_{i,j}^k d_{i,j}\Big) 
    \Bigg\}, ~\forall k,
\eeqn
where $\mathcal{F}^k$ is the feasible set of $(x^k, y^k)$ satisfying constraints $(\ref{eef1})-(\ref{eef71})$ of the follower problem for service $k$. 
The platform aims to maximize its profit
by jointly optimizing the EN activation, service placement, and resource pricing decisions.
Note that since the objective of the platform is profit maximization,
the inequalities in (\ref{eq5})  become equalities at the optimality. Thus, we can rewrite (\ref{ecost}) as: $ Cost_j^{\sf e} = c_j z_j + q_j  \frac{\sum_k y_j^k}{C_j},~ \forall j$. The profit in (\ref{profiteq}) can also be rewritten as in (\ref{objlead}).
The follower problems (i.e., the lower-level problems) 
serve as  constraints of the leader problem, as shown in (\ref{fol}).

\section{Solution Approach}
\label{sol}

The bilevel program (\ref{objlead})-(\ref{fol}) is generally hard to solve due to not only the constraints (\ref{fol}) in forms of  optimization problems but also the bilinear terms $p_j y_j^k$ in the objective function (\ref{objlead}). 
For the special case with a single EN, we can solve the bilevel problem analytically since the number of possible resource prices (i.e., $V$) at the EN is finite and small. 
Please refer to \textit{Appendix \ref{1EN}} for the detailed solution.

In the following, we tackle  the general case with multiple ENs. First, we  present the KKT-based approach to reformulate the   bilevel problem
into an
equivalent single-level MILP.
Specifically, the bilevel program is transformed into an MPEC 
by replacing each follower problem by its KKT conditions.  Then, by combining several linearization approaches and the strong duality theorem \cite{sgab12,dual}, the resulting MPEC can be recast as an MILP. 
Instead of relying on the KKT conditions, the second approach employs LP duality to convert the bilevel problem into an equivalent MILP with considerably less number of constraints and integer variables compared to the one obtained from the KKT-based approach.  
\subsection{KKT-based Reformulation}
First, recall that  the optimization variables $p$ and $t$ of the leader problem are parameters of the follower problems. Thus, for fixed values of $p$ and $t^k$,   the  lower-level  problem (\ref{objf})-(\ref{eef71})  is a linear program, and thus convex. As a result, the KKT conditions are necessary
and sufficient for optimality. Consequently, we can replace each follower problem by its corresponding KKT conditions, including the stationary, complementary slackness, primal feasibility, and dual feasibility conditions \cite{boyd}. The primal feasibility conditions are (\ref{eef1})--(\ref{eef71}). The Lagrangian of the follower problem (\ref{objf})--(\ref{eef71}) is: 
\beqn
\label{lagranian}
L^k (x^k, y^k,  d^{k, \sf a}, \xi^k, \sigma^k, \tau^k, \mu_1^k, \lambda^k, \Gamma^k, \eta^k, \mu_2^k, \zeta^k, \epsilon^k)   \\ \nonumber =   
\sum_j p_j  y_j^k  + w^k \Big(\sum_i x_{i,0}^k d_{i,0} + \sum_{i,j} x_{i,j}^k d_{i,j}\Big) + p_0 y_0^k
\\ \nonumber + \sum_j \lambda_j^k \Big(\sum_i x_{i,j}^k  - y_j^k \Big)   +  \mu_1^k \Big(\sum_i x_{i,0}^k - y_0^k \Big)  \\ \nonumber +  \sum_i \xi_i^k \Big(  R_i^k - x_{i,0}^k   - \sum_j x_{i,j}^k   \Big)  + \sum_j \Gamma_j^k \Big( y_j^k - C_j t_j^k \Big) 
 \\ \nonumber + \sum_{i,j} \eta_{i,j}^k \Big( x_{i,j}^k - a_{i,j}^k R_i^k \Big) 
   + \sum_i \tau_i^k \Big( d_{i}^{k, \sf a} - D^{k, \sf m} \Big) 
   \\ \nonumber + \sum_i \sigma_i^k \Big( x_{i,0}^k d_{i,0} + \sum_j x_{i,j}^k d_{i,j}  - d_{i}^{k, \sf a} R_i^k \Big) 
 - \sum_i x_{i,0}^k \zeta_{i,0}^k \\ \nonumber + \mu_2^k \Big( p_0 y_0^k + \sum_j p_j y_j^k - B^k \Big)  - \sum_{i,j} x_{i,j}^k \epsilon_{i,j}^k.
\eeqn
Thus, the KKT stationary conditions are given as follows:
\beqn
\label{kttx0}
\frac{\delta L}{\delta x_{i,0}^k} = w^k d_{i,0} - \xi_i^k + \sigma_i^k d_{i,0} + \mu_1^k - \zeta_{i,0}^k = 0, ~ \forall i,k\\
\label{kttx}
\frac{\delta L}{\delta x_{i,j}^k} = w^k d_{i,j} - \xi_i^k + \sigma_i^k d_{i,j} + \lambda_j^k  \nonumber \\ + \eta_{i,j}^k - \epsilon_{i,j}^k = 0, ~ \forall i,j,k \\
\label{kktda}
\frac{\delta L}{\delta d_{i}^{k, \sf a}}   = - \sigma_i^k R_i^k + \tau_i^k  = 0, \quad \forall i,k.\\
\label{kkty0}
\frac{\delta L}{\delta y_0^k}   = p_0  - \mu_1^k + p_0 \mu_2^k = 0, \quad \forall k \\
\label{kkty}
\frac{\delta L}{\delta y_j^k}   = p_j - \lambda_j^k + \Gamma_j^k + p_j \mu_2^k = 0, \quad \forall j,k 
\eeqn
Also, the complementary slackness, dual feasibility, and the primal feasibility conditions of the follower problems render:
\beqn
\label{eq:cc1}
0 \leq \tau_i^k ~\bot ~  D^{k, \sf m} - d_{i}^{k, \sf a}  \geq 0,~ \forall i,k \\
\label{eq:cc2}
0 \leq \mu_1^k \bot ~ y_0^k - \sum_i x_{i,0}^k \geq 0, \forall k  \\
\label{eq:cc3}
0 \leq \lambda_j^k ~\bot~ y_j^k - \sum_i x_{i,j}^k   \geq 0, ~\forall j,k \\
\label{eq:cc4}
0 \leq \Gamma_j^k ~\bot~ C_j t_j^k - y_j^k \geq 0,~\forall j,k \\
\label{eq:cc5}
0 \leq \eta_{i,j}^k ~\bot~ a_{i,j}^k R_i^k - x_{i,j}^k \geq 0,~\forall i,j,k \\
\label{eq:cc6}
0 \leq \mu_2^k ~\bot~ B^k - p_0 y_0^k - \sum_j p_j y_j^k \geq 0,~\forall k \\
\label{eq:cc7}
0 \leq \zeta_{i,0}^k ~\bot~ x_{i,0}^k \geq 0,~\forall i,k\\
\label{eq:cc8}
0 \leq \epsilon_{i,j}^k ~\bot~ x_{i,j}^k \geq 0,~\forall i,j,k.
\eeqn
Note that 
$0 \leq a ~\bot~ b \geq 0$ means $a \geq 0, ~b \geq 0$, and $a b = 0$. Constraints (\ref{eq:cc1})-(\ref{eq:cc8}) are called complementarity constraints or equilibrium constraints.
By replacing constraints  (\ref{fol}) for the follower problems  
with the set of constraints (\ref{eef1})--(\ref{eef71}) and (\ref{kttx0})--(\ref{eq:cc8}), 
the bilevel program
(\ref{objlead})-(\ref{fol}) becomes an MPEC problem. 
This MPEC problem has three sources of nonlinearity, including: i) the complementarity constraints (\ref{eq:cc1})-(\ref{eq:cc8});  ii)  the bilinear terms $p_j \mu_2^k$ in  (\ref{kkty}); and iii) the bilinear term $\sum_{j,k} p_j  y_j^k$ in the objective function (\ref{objlead}).
To convert the MPEC problem (i.e., an MINLP) into an MILP, we need to linearize these nonlinear terms.

First, the nonlinear complementarity constraints (\ref{eq:cc1})--(\ref{eq:cc8}) can be transformed into equivalent exact linear constraints by using the Fortuny-Amat transformation \cite{bigM}.
Specifically, the complementarity condition $0 \leq a ~\bot~ b \geq 0$ 
is equivalent to the following set of mixed-integer linear constraints:
\beqn
a \leq (1 - u) M_0;~ b \leq u M_0;~ 
a \geq 0;~ b \geq 0;~ u \in \{0,~1 \},
\eeqn
where $M_0$ is a sufficiently large constant, , which is often referred to as ``bigM''.
Therefore, the set of constraints (\ref{eq:cc1})--(\ref{eq:cc8}) can be rewritten as:
 \beqn
\label{eq:coms}
D^{k, \sf m} - d_{i}^{k, \sf a} \leq \psi_i^k M_1;~ \tau_i^k \leq (1-\psi_i^k) M_1,~~\forall i,k  \\
y_0^k - \sum_i x_{i,0}^k \leq v_1^k M_2;~ \mu_1^k \leq (1-v_1^k)M_2,~~\forall k  \\
y_j^k - \sum_i x_{i,j}^k \leq \kappa_j^k M_3;~ \lambda_j^k \leq (1-\kappa_j^k) M_3,~~\forall j,k  \\ 
C_j t_j^k - y_j^k \leq \theta_j^k M_4;~ \Gamma_j^k \leq (1-\theta_j^k)M_4,~~\forall j,k  \\ 
a_{i,j}^k R_i^k - x_{i,j}^k \leq \rho_{i,j}^k M_5;~ \eta_{i,j}^k \leq (1-\rho_{i,j}^k) M_5,~\forall i,j,k  \\ 
B^k - p_0 y_0^k - \sum_j p_j y_j^k \leq v_2^k M_6;~ \mu_2^k \leq (1-v_2^k)M_6,~\forall k  \\
x_{i,0}^k \geq 0;~x_{i,0}^k \leq \Phi_{i,0}^k M_7;~ \zeta_{i,0}^k \leq (1- \Phi_{i,0}^k) M_7,~\forall i,k  \\
x_{i,j}^k \geq 0;~x_{i,j}^k \leq \Omega_{i,j}^k M_8;~ \epsilon_{i,j}^k \leq (1- \Omega_{i,j}^k) M_8,~\forall i,j,k  \\
 \tau_i^k \geq 0 ;~\mu_1^k \geq 0 ;~ \lambda_j^k \geq 0 ;~\Gamma_j^k \geq 0 \\
\eta_{i,j}^k \geq 0 ;~\mu_2^k \geq 0 ;~ \zeta_{i,0}^k \geq 0 ;~ \epsilon_{i,j}^k \geq 0 \\
\label{eq:come}
\psi_i^k,~ \kappa_j^k, ~\theta_j^k, ~\rho_{i,j}^k, ~\Phi_{i,0}^k, ~\Omega_{i,j}^k, ~v_1^k,~ v_2^k \in \{0,1\},~\forall i,j,k,
\label{eq:comes}
\eeqn
where $M_1$, $M_2$, $M_3$, $M_4$, $M_5$, $M_6$, $M_7$ and $M_8$ are sufficiently large numbers.
 For the bilinear terms $p_j \mu_2^k$, using (\ref{leaderc4}), we can rewrite it as:
\beqn
\label{pmu2linear}
p_j \mu_2^k = \sum_v p_j^v r_j^v \mu_2^k = \sum_v p_j^v \pi_j^{v,k},
\eeqn
where $\pi_j^{v,k} = r_j^v \mu_2^k$. Note that $\pi_j^{v,k}$ is a continuous variable and we have $\pi_j^{v,k} = \mu_2^k$ if $r_j^v = 1$ and $\pi_j^{v,k} = 0$, otherwise. 
Hence, using (\ref{pmu2linear}), the bilinear term $p_j \mu_2^k$ can be written as a linear function of $\pi_j^{k} = (\pi_j^{1,k}, \ldots, \pi_j^{V,k})$. 
Additionally, the constraints $\pi_j^{v,k} = r_j^v \mu_2^k, \forall j,k$ can be implemented through the following linear inequalities \cite{linear2}:
\beqn
\label{pmulinear1}
\pi_j^{v,k} \leq M r_j^v,~~ \forall j, k, v;~~
\pi_j^{v,k} \leq \mu_2^k,~~ \forall j, k, v\\
\label{pmulinear2}
\pi_j^{v,k} \geq 0, ~~ \forall j, k, v;~~ \pi_j^{v,k} \geq \mu_2^k + M r_j^v - M, ~~ \forall j, k, v,
\eeqn
where $M$ is a sufficiently large number.

We assume that the bilevel problem has an optimal solution and the strong duality theorem holds for every follower problem. Then, the strong duality theorem gives us the following:
\beqn
\label{pylinear}
 \sum_j p_j y_j^k  + p_0 y_0^k + w^k \Big(\sum_i x_{i,0}^k d_{i,0} + \sum_{i,j} x_{i,j}^k d_{i,j}\Big) = \nonumber \\  - \sum_{i,j} R_i^k a_{i,j}^k \eta_{i,j}^k  - \sum_j C_j t_j^k \Gamma_j^k - B^k \mu_2^k  \nonumber \\ + \sum_i R_i^k  \xi_i^k  - \sum_i D^{k, \sf m} \tau_i^k, ~~\forall k.
\eeqn
Hence, using (\ref{pylinear}), the bilinear term $\sum_{j,k} p_j  y_j^k$ 
can be written as the sum of several linear terms. Note that the bilinear terms $t_j^k \Gamma_j^k$ in  (\ref{pylinear}) is a product of a continuous variable and a binary variable. Therefore, we can linearize it similar to what we did for the bilinear terms $ r_j^v \mu_2^k$ using (\ref{pmulinear1}) and (\ref{pmulinear2}).

Based on the linearization steps described above, we can then express the bilevel problem (\ref{objlead})-(\ref{fol}) with an equivalent single-level MILP as follows:
\beqn
\label{objlead2}
(\mathcal{P}_1): \underset{p,z,t,x,y}{\text{max}}  ~ \textit{Rev} -  \Bigg[ \sum_j  \big(c_j z_j + q_j  \frac{\sum_k y_j^k}{C_j}\big)
+ \sum_{j,k}  \phi_j^k t_j^k  \Bigg] \nonumber
\eeqn
subject to
\beqn
Rev = - \sum_k \Bigg[ p_0 y_0^k + w^k \Big(\sum_i x_{i,0}^k d_{i,0} + \sum_{i,j} x_{i,j}^k d_{i,j}\Big) \nonumber \\   + \sum_{i,j} R_i^k a_{i,j}^k \eta_{i,j}^k  + \sum_j C_j t_j^k \Gamma_j^k   +  B^k \mu_2^k \nonumber \\  + \sum_i R_i^k  \xi_i^k  + \sum_i D^{k, \sf m} \tau_i^k \Bigg] \nonumber \\
p_j - \lambda_j^k + \Gamma_j^k + \sum_v p_j^v \pi_j^{v,k} = 0, ~\forall j,k ;~(\ref{pmulinear1}), ~(\ref{pmulinear2}) \nonumber \\
  (\ref{eef1})-(\ref{eef71}), ~(\ref{leaderc0}) - (\ref{leaderc4}), ~(\ref{kttx0})-(\ref{kkty0}), ~(\ref{eq:coms})-(\ref{eq:come}),\nonumber
\eeqn
where $\textit{Rev}$ is the revenue of the platform from selling edge resources, i.e., $\textit{Rev} = \sum_{j,k} p_j y_j^k$. Problem   ($\mathcal{P}_1$) is a large-scale MILP, which can be solved by MILP solvers.


\subsection{Duality-based Reformulation}
Instead of using KKT conditions, we can utilize the LP duality to transform the original bilevel problem into an equivalent MILP. 
We  first write the dual maximization form of each lower-level minimization problem (\ref{objf})--(\ref{eef71}). Subsequently, we can replace each lower-level problem by its corresponding 
dual feasibility conditions, as well as equating the primal and dual objective functions \cite{boyd}.
The dual problem  of the follower problem 
(\ref{objf})--(\ref{eef71}) of  service $k$ is given below: 
\beqn
\label{objdf}
\underset{\xi^k,~\sigma^k,~\tau^k,~\mu_1^k,~\mu_2^k,~\lambda^k,~\eta^k,~\Gamma^k}{\text{maximize}} ~ \sum_i R_i^k  \xi_i^k - B^k \mu_2^k  \nonumber \\ - \sum_i \sum_j R_i^k a_{i,j}^k \eta_{i,j}^k - \sum_j C_j t_j^k \Gamma_j^k - \sum_i D^{k, \sf m} \tau_i^k  
\eeqn
~~subject to 
\beqn
\label{eedf1}
      \lambda_j^k - \Gamma_j^k \leq \ p_j (1 + \mu_2^k), ~~ \forall j \\
\label{eedf2}
      \mu_1^k \leq  p_0 (1+ \mu_2^k) \\
\label{eedf3}
      - R_i^k \sigma_i^k - \tau_i^k \leq 0, ~~\forall i \\
\label{eedf4}
     \xi_i^k + \sigma_i^k d_{i,j} - \lambda_j^k - \eta_{i,j}^k + \epsilon_{i,j}^k \leq w^k d_{i,j}, ~~\forall i,j \\
\label{eedf5}
    \xi_i^k + \sigma_i^k d_{i,0} - \mu_1^k + \zeta_{i,0}^k \leq w^k d_{i,0}, ~~ \forall i \\
\label{eedf6} 
 \eta_{i,j}^k \geq 0, ~\forall i,j;~~ \mu_1^k \geq 0;~~\mu_2^k \geq 0 \\
 \label{eedf61} 
 \tau_i^k \geq 0, ~\forall i; \quad \lambda_j^k \geq 0, ~\forall j; ~~ \Gamma_j^k \geq 0,~\forall j.
\eeqn
The dual feasibility constraints are (\ref{eedf1})-(\ref{eedf61}).
Thus, the complete form of the final MILP optimization problem resulting from the duality-based reformulation is given as follows:

\beqn
(\mathcal{P}_2): ~\underset{p,x,y,z,t,\lambda,\xi,\eta,\tau,\mu_1,\mu_2,\Gamma,\sigma}{\text{maximize}} ~ \sum_k \textit{Rev}^k  \\ \nonumber-  \Bigg[ \sum_j  \big(c_j z_j + q_j  \frac{\sum_k y_j^k}{C_j}\big)
+ \sum_{j,k}  \phi_j^k t_j^k  \Bigg]
\eeqn
subject to 
\beqn
\label{revk}
\textit{Rev}^k   + p_0 y_0^k + w^k \Big(\sum_i x_{i,0}^k d_{i,0} +\sum_{i,j} x_{i,j}^k d_{i,j}\Big) = \nonumber \\  - \sum_{i,j} R_i^k a_{i,j}^k \eta_{i,j}^k  - \sum_j C_j t_j^k \Gamma_j^k - B^k \mu_2^k \nonumber \\ + \sum_i R_i^k  \xi_i^k   - \sum_i D^{k, \sf m} \tau_i^k,~~\forall k   \\
 (\ref{eedf1})-(\ref{eedf61}), ~~\forall k \\
 (\ref{eef1})-(\ref{eef71}), ~(\ref{leaderc0}) - (\ref{leaderc4}). \nonumber 
\eeqn
Note that $\textit{Rev}^k$
is the revenue from selling edge resources to service $k$, i.e.,  $\textit{Rev}^k = \sum_j p_j y_j^k,~\forall k$. 
Constraints (\ref{revk}) in problem ($\mathcal{P}_2$) enforce the primal objective function equals the dual objective function, which indeed expresses the strong duality theorem. We can linearize the bilinear terms $t_j^k \Gamma_j^k$ in (\ref{revk}) by applying the same procedure that we employed in the KKT-based transformation approach.
Consequently, using (\ref{revk}), we can linearize the bilinear terms $p_j y_j^k$. 
Finally, the dual feasibility constraints (\ref{eedf1})-(\ref{eedf61}) should hold for all $k$. 

Compared to the MILP problem ($\mathcal{P}_1$), it is easy to see that the MILP problem ($\mathcal{P}_2$) does not need to deal with the 
complementarity constraints (\ref{eq:cc1})-(\ref{eq:cc8}) or their equivalent linear constraints
(\ref{eq:coms})-(\ref{eq:come}). Thus, it drastically reduces the number of constraints and auxiliary binary variables in problem ($\mathcal{P}_1$). 
Table \ref{table2} compares the number of constraints and binary/continuous variables in ($\mathcal{P}_1$) and ($\mathcal{P}_2$).



\begin{table*}[ht]
\centering
\begin{tabular}{ |c|c|c|}
 \hline
 \  & KKT-based reformulation ($\mathcal{P}_1)$ & Duality-based reformulation ($\mathcal{P}_2)$ \\
  \hline
  \# constraints  
                 & $1+4N+9K+2NK(4V+7)+3MK(3N+4)$  & $2(2N+3K)+ 4NK(M+2V) + K(7M+11N)$  \\
 \hline
  \# binary variables  
                  &$N(K+V+1) + 2K(M+1)(N+1)$ & $ N(K+V+1)$ \\
 \hline
 \# continuous variables 
                   &$K[3(MN+M+1) + 2N(V+2)]$ & $K[3 + 2M(N+1) + 2N(V+2)]$ \\
\hline
\end{tabular}
 \caption{Problem size comparison between ($\mathcal{P}_1$) and ($\mathcal{P}_2$) }
 \label{table2}
\end{table*}

Computational complexities: Bilevel optimization is notoriously hard to solve. It is well-known that even the simplest  linear bilevel program, where both the leader and follower problems are linear programs, is strongly NP-hard \cite{bilevel1,bilevel2,bilevel3, bilevel4}. Furthermore, checking strict local optimality in linear bilevel programming is also NP-hard \cite{bilevel4}. In this section, we have shown how to  convert the formulated MINBP into an MILP.  
Note that integer linear program (ILP) is NP-hard \cite{mipbook2, kannan78, book, np1, np3}. Indeed, ILP is NP-complete if we consider it as a decision problem \cite{book, np1, np3}. Since ILP is a special case of MILP, MILP is at least as hard as ILP, and thus,  is also NP-hard. In general, most MILP problem instances are hard to solve, except for some special kinds of ILP (e.g., an ILP instance where the constraint matrix is a totally unimodular matrix). In this work, we do not aim to propose efficient approximation algorithms to solve the formulated bilevel program. 
Instead, our goal is to transform the original challenging MINBP into an MILP, which can be solved efficiently and exactly by modern MILP solvers. Most of the current MILP solvers implement the branch-and-bound (BnB) method and the branch-and-cut (BnC) method \cite{book} to solve MILP instances. The computational time of these methods generally depends on the problem size, especially the number of integer variables \cite{book}. Hence, from Table \ref{table2}, it is easy to see that solving ($\mathcal{P}_2$) is typically 
faster than solving ($\mathcal{P}_1$). Also,  modern MILP solvers normally employ proprietary heuristic algorithms to  reduce the problem size in a pre-solve phase before applying BnB and/or BnC methods. Thus, the solvers  can efficiently solve large-scale MILP instances.

\section{Numerical Results}
\label{sim}

\subsection{Simulation Setting}
Similar to the previous work \cite{mjia17,duongton,duongiot}, 
we adopt the widely-used Barabasi-Albert model \cite{ralb99} with an attachment rate of 2 to generate an edge network topology with 100 nodes. 
The link
delay between each pair of nodes is randomly generated in the range of [2,~5] \textit{ms}. The network delay between any two nodes
is computed as the delay of the shortest path between them. 
In the \textbf{base case} scenario, we consider a small system with 6 services, 10 APs and 4 ENs, which are picked randomly in the set of 100 nodes. Thus, in the base case, $M$ = 10, $N$ = 4, $K$ = 6. 
We will also run the proposed algorithms on different system sizes  for sensitivity analysis later. 
The delay between each AP and the remote cloud is set to be 60 \textit{ms}. The maximum delay threshold $D^{k, \sf m}$ for each service is selected randomly between 30 \textit{ms} and 100 \textit{ms}.
During the scheduling horizon, for each service, the resource demand (i.e., workload) in each area is randomly drawn in the range of [20, 35] vCPUs. 


Each EN is chosen randomly from the set of Amazon EC2 M5 instances. 
Using the hourly price of a general purpose m5d.xlarge Amazon EC2 instance \cite{EC2} as  reference, the unit resource price at the cloud is set to be 0.01 $\$$/vCPU while 
the set of possible unit prices of edge resources is [0.01, 0.02, 0.03, 0.04, 0.05] $\$$/vCPU. The fixed and variable operational costs $c_j$ and $q_j$ of each EN $j$ are set in the range of [\$0.05, \$1.8] and [\$0.04, \$1.44], respectively, depending on the size of the EN. The delay penalty parameters $w^k$ are generated randomly in the range of [$10^{-5}$, $10^{-3}$] $\$/$(vCPU.ms). Additionally, the placement cost of  each service at each EN is set to be \$0.02 (i.e., $\phi_j^k = 0.02,~\forall j,k$). The size of each service is randomly generated between 10GB and 100GB. The budget of each service is chosen between \$150 and \$300. 

Unless otherwise stated, the default setting is used in most experiments. 
Our computational study is made through Matlab 2020b software and solved by Gurobi solver on an Intel Core i7-10510U CPU  and 16 GB RAM laptop.
The computational time limit is set to 10,000 seconds.

\subsection{Performance Evaluation}

\subsubsection{Comparison between the KKT-based and duality-based reformulation approaches}
We  compare  the computational time between the KKT-based and duality-based solution methods. Both methods allow us to compute an optimal solution to the original bilevel problem.  The two methods are compared under different system sizes by varying the numbers of APs, ENs, and services. The  computational results 
are reported in Table \ref{table1}. Note that ``NA'' implies a method cannot produce a solution within the time limit. 
As expected, the duality-based approach offers superior performance compared to the KKT-based approach. It is because of the smaller size of the MILP obtained from the duality-based method 
compared to the one obtained from the KKT-based method. Another disadvantage of the KKT-based method is that we have to choose suitable bigM values, which
 greatly affects its running time. 
We thus adopt the duality-based method  to generate results in the following experiments.
\begin{table}[ht]
\begin{tabular}{ |p{2.5cm}|p{2.5cm}|p{2.5cm}|}
 \hline
 \multicolumn{3}{|c|}{$K$ = 6, $N$ = 4, varying $M$} \\
 \hline
  $M$ & Duality (seconds) & KKT (seconds)\\
 \hline
 2   & 5.8930    &8.8937\\
 4 &7.9730 &25.7110\\
 6 &6.9494 &99.5641\\
 \hline
 \multicolumn{3}{|c|}{$K$ = 6, $M$ = 10, varying $N$} \\
 \hline
  $N$ & Duality (seconds) & KKT (seconds)\\
 \hline
 4   &11.993     &53.4655\\
 6   &175.0676   &515.0104\\
 8   &288.42     &4919.8 \\ 
 \hline
 \multicolumn{3}{|c|}{$M$ = 10, $N$ = 4, varying $K$} \\
 \hline
  $K$ & Duality  (seconds) &KKT  (seconds)\\
 \hline
 4   &3.9131   &39.5804\\
 6   &9.8309   &53.5309\\
 8   &74.3804  & NA \\ 
 10  &146.3724 & NA\\
 \hline
\end{tabular}
 \caption{Computational time comparison between the duality-based and KKT-based reformulation methods}
 \label{table1}
\end{table}

\subsubsection{Comparison between dynamic, flat, and average pricing schemes} 
\begin{figure}[h!]
      \subfigure[Varying cloud price]{\includegraphics[width=0.245\textwidth,height=0.11\textheight]{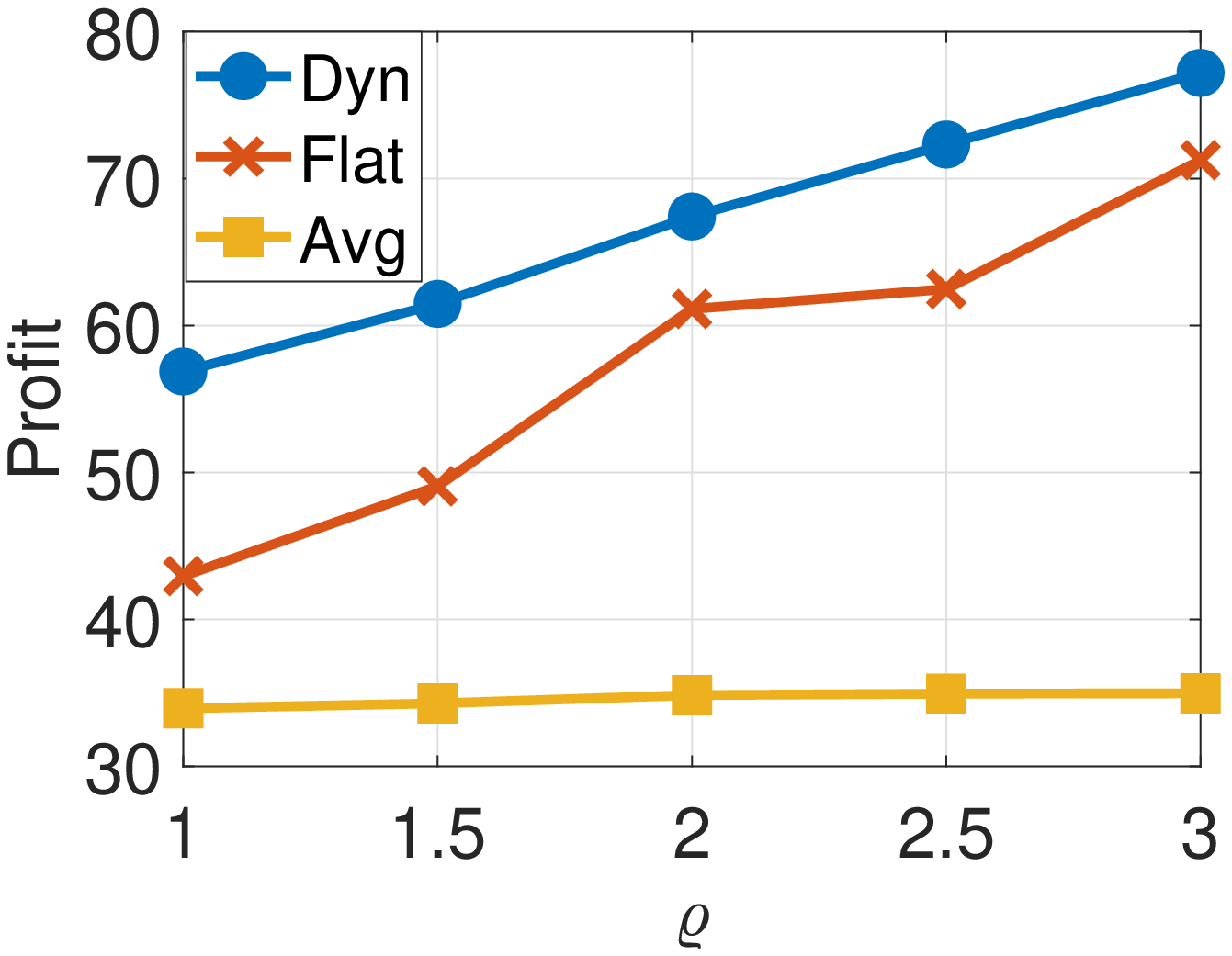}
       \label{fig:Profit varying cc}}  \hspace*{-2.1em}
     \subfigure[Varying delay penalty]{\includegraphics[width=0.245\textwidth,height=0.11\textheight]{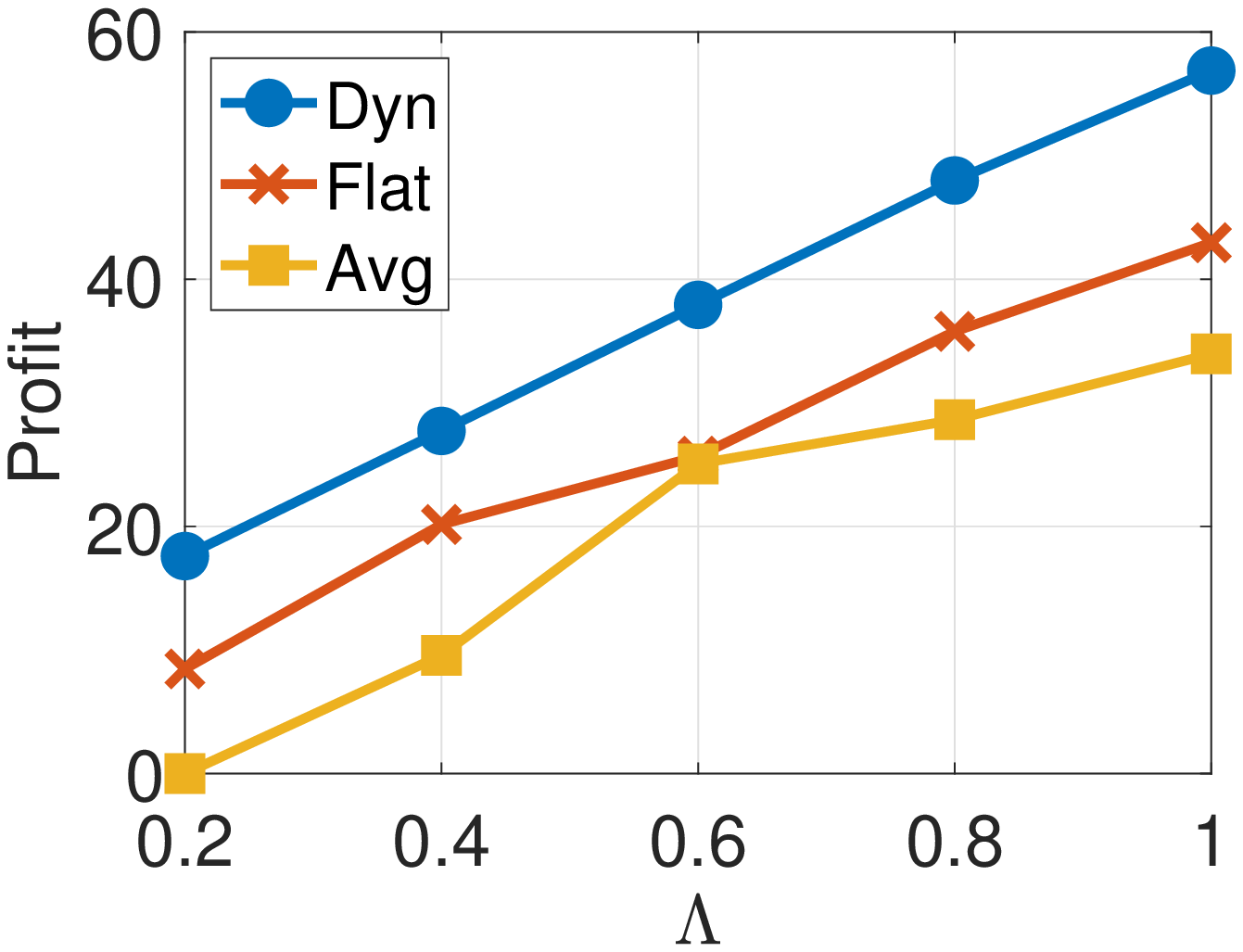}
       \label{fig:Profit varying w}}
      \vspace{-0.2cm}
      \caption{Performance comparison between \textit{Dyn}, \textit{Flat}, and \textit{Avg}}
      \label{fig:Comparison DFA}
\end{figure}

The dynamic pricing  scheme (\textit{Dyn}) is the proposed model where the resource prices at the ENs can be different to balance supply and demand. In the flat pricing  scheme (\textit{Flat}), we solve the same bilevel model with an extra constraint in the leader problem that enforces the resource prices at all the ENs to be equal. In the average pricing  scheme (\textit{Avg}), the unit resource prices at the ENs are the same, which are simply set to the average value in the set of possible prices. 
We first examine the impact of cloud resource price on the profit of the platform, as shown in Fig. \ref{fig:Profit varying cc}. We define $\varrho$ as the scaling factor of the cloud price. For example, when $\varrho = 1.5$, the cloud resource price in the base case increases $1.5$ times.   

It can be seen 
the proposed dynamic pricing scheme significantly outperforms the 
other schemes. 
Indeed, with  dynamic pricing, the platform can reduce the prices of under-demanded ENs, which  incentivizes  the services to reallocate their workload to these nodes, thus, improving its revenue and  resource utilization of these nodes. Also,  the prices of highly-demanded ENs are often set to the maximum price value. 
Thus, the profit in \textit{Dyn} is highest due to higher edge resource demand and generally higher prices. Since the service buys more edge resources, the average network delay of each service tends to decrease.  
In both the flat and average pricing schemes, the prices at all the ENs are the same. Hence, the services do not have any incentive for load shifting. 
Furthermore, we can see that the profits in all these schemes increase as the cloud resource price increases. This is because an increase in the cloud resource price would encourage the services to shift more workload to the ENs. Therefore, the platform can sell more edge resources and increase its revenue and profit.  

Fig. \ref{fig:Profit varying w} shows the impact of  delay penalty on the profit of the platform. Let   $\Lambda$ be the scaling factor of the delay penalty parameters $w^k$. It is easy to see the superior performance of the dynamic pricing scheme compared to the flat and average pricing schemes. 
Also, the profit increases as the delay penalty parameters increase. 
It is because when the services are more delay-sensitive, they are willing to pay more for edge resources to reduce the overall delay for their users. Hence, the platform can increase the edge resource prices to increase its profit.

\begin{figure}[hbt!]
	\centering
		\includegraphics[width=0.3\textwidth,height=0.13\textheight]{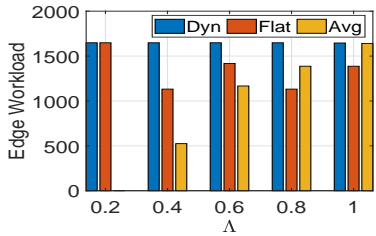}
			\caption{Total workload allocated at the edge}
	 \label{fig:Edge Traffic varying w}
\end{figure}


Fig.~\ref{fig:Edge Traffic varying w} depicts the total workload at the edge under the three pricing models with varying delay penalty. 
In \textit{Dyn}, all the demands are served  at the edge and there is no cloud traffic. The reason is that 
\textit{Dyn} allows the platform to  adjust  edge prices. Hence, the prices at under-demanded ENs can be reduced so that the gain from offloading at these nodes outweighs the price difference between the cloud and each node. In \textit{Flat}, the platform is less flexible in setting the prices since the prices at all the ENs are equal.
A low edge resource price will affect the platform's revenue. To maximize the profit, the flat price should not be too low. 
Thus, in \textit{Flat}, a portion of demand will go to the cloud.  
 Finally, in the average pricing scheme, the prices at the ENs are the same and fixed. Thus, when  the delay penalty increases, the gain from offloading to the edge increases and less workload goes to the cloud.


\begin{figure}[h!]
     \subfigure[Varying $M$ and resource demand]{\includegraphics[width=0.245\textwidth,height=0.11\textheight]{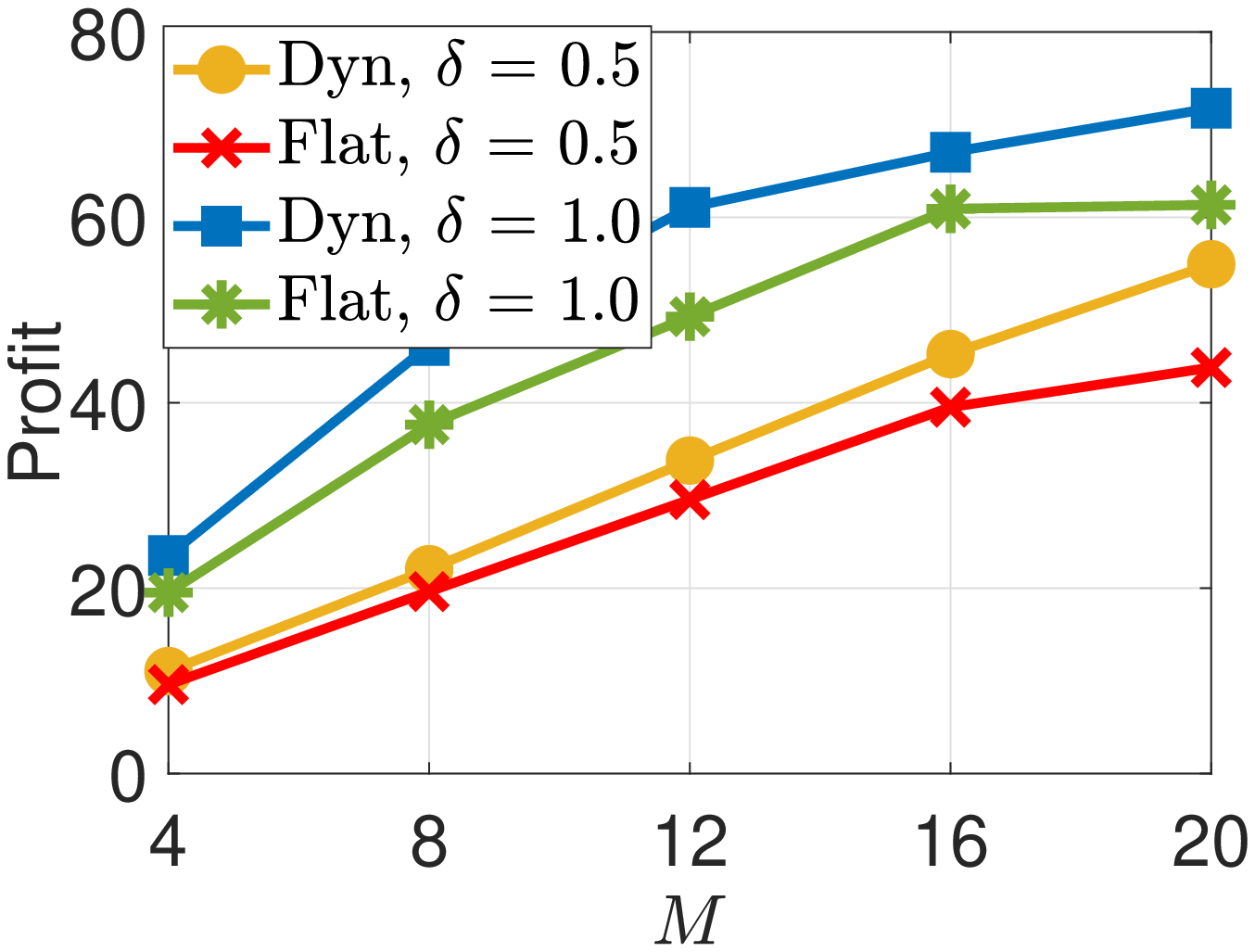}
     \label{fig:M varying R}}  \hspace*{-1.9em}
     \subfigure[Varying $M$ and delay penalty]{\includegraphics[width=0.245\textwidth,height=0.11\textheight]{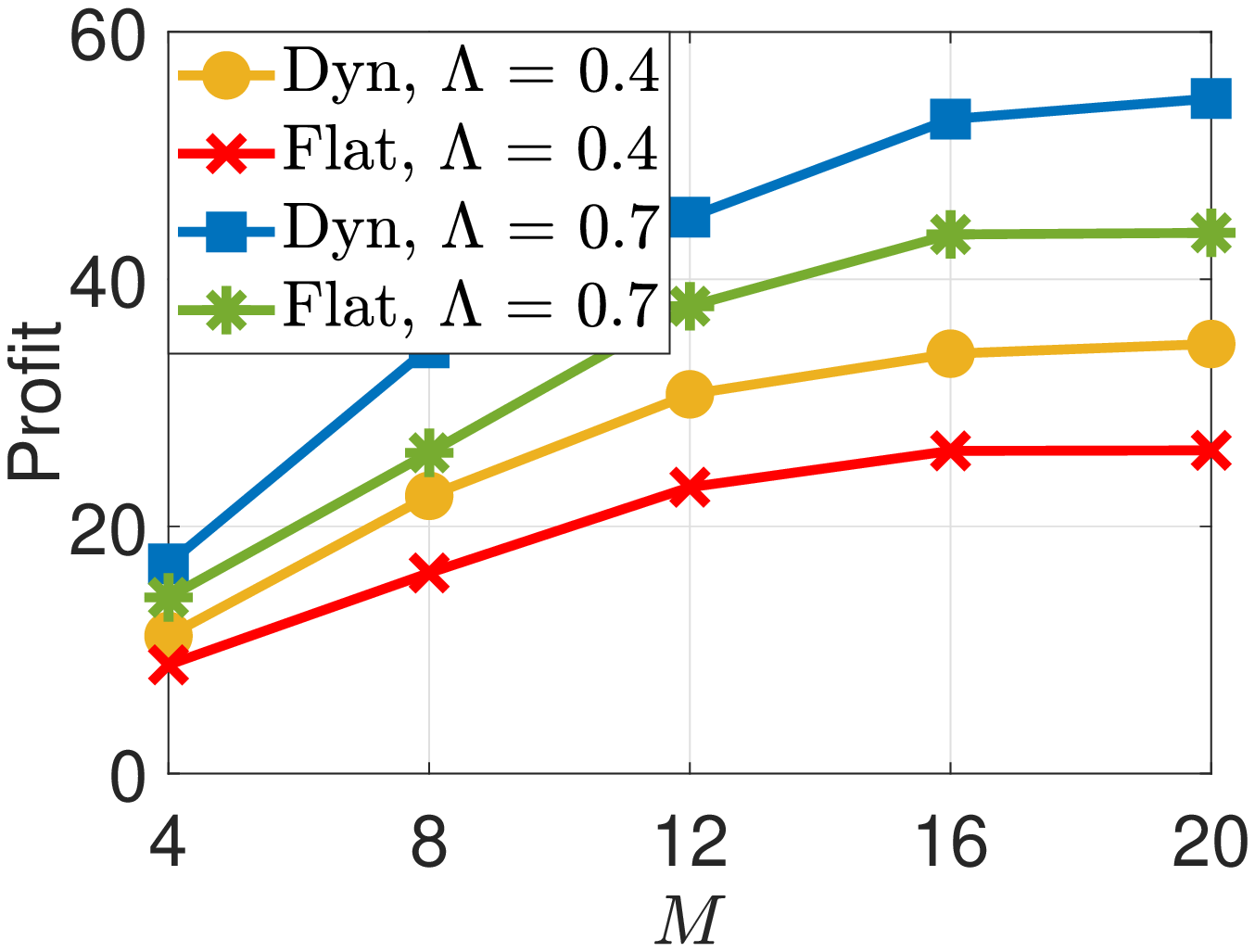}
     \label{fig:M varying w}}  
     \subfigure[Varying $M$ and EN capacities]{\includegraphics[width=0.245\textwidth,height=0.11\textheight]{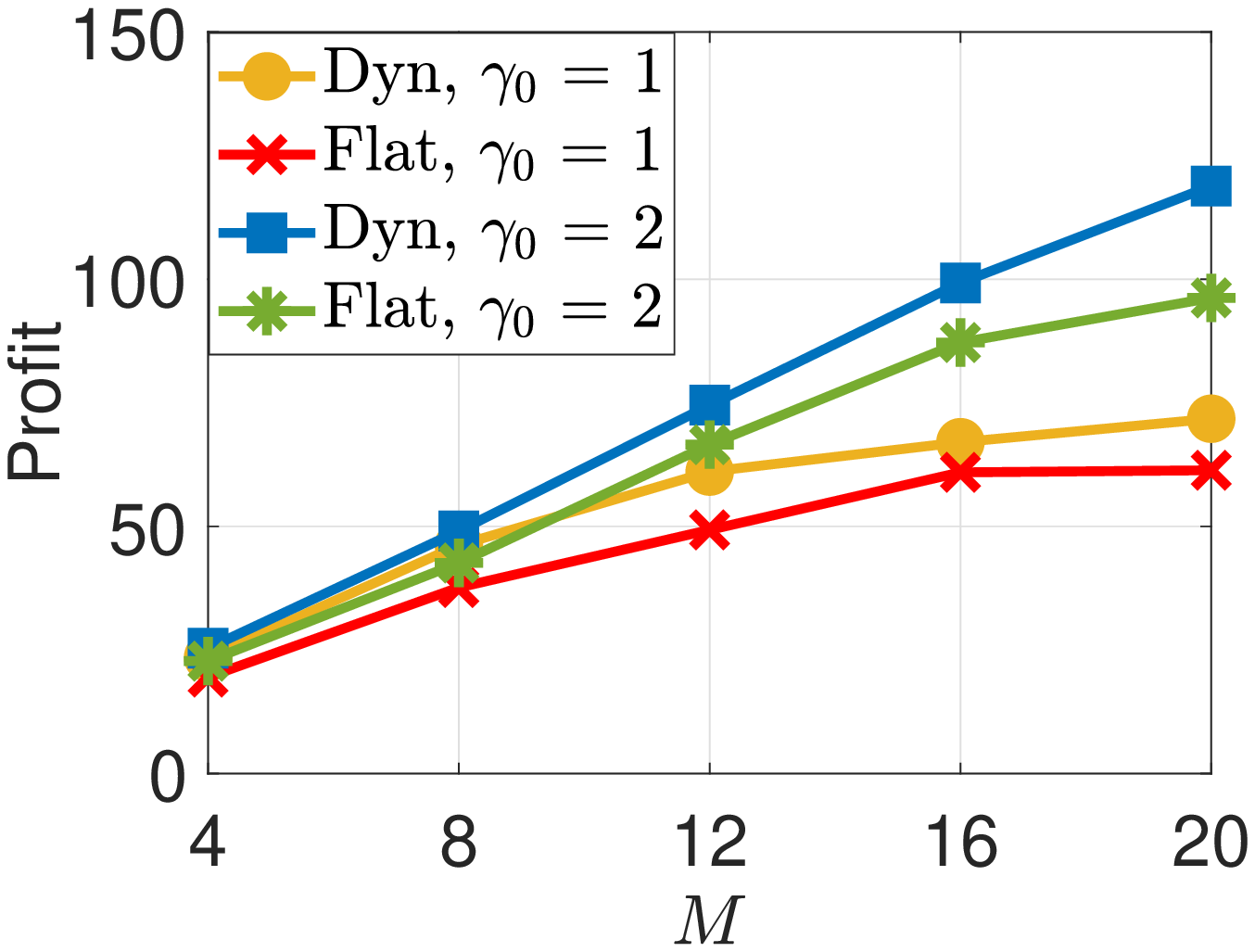}
     \label{fig:M varying C}} \hspace*{-1.9em}
     \subfigure[Varying $M$ and cloud price ]{\includegraphics[width=0.245\textwidth,height=0.11\textheight]{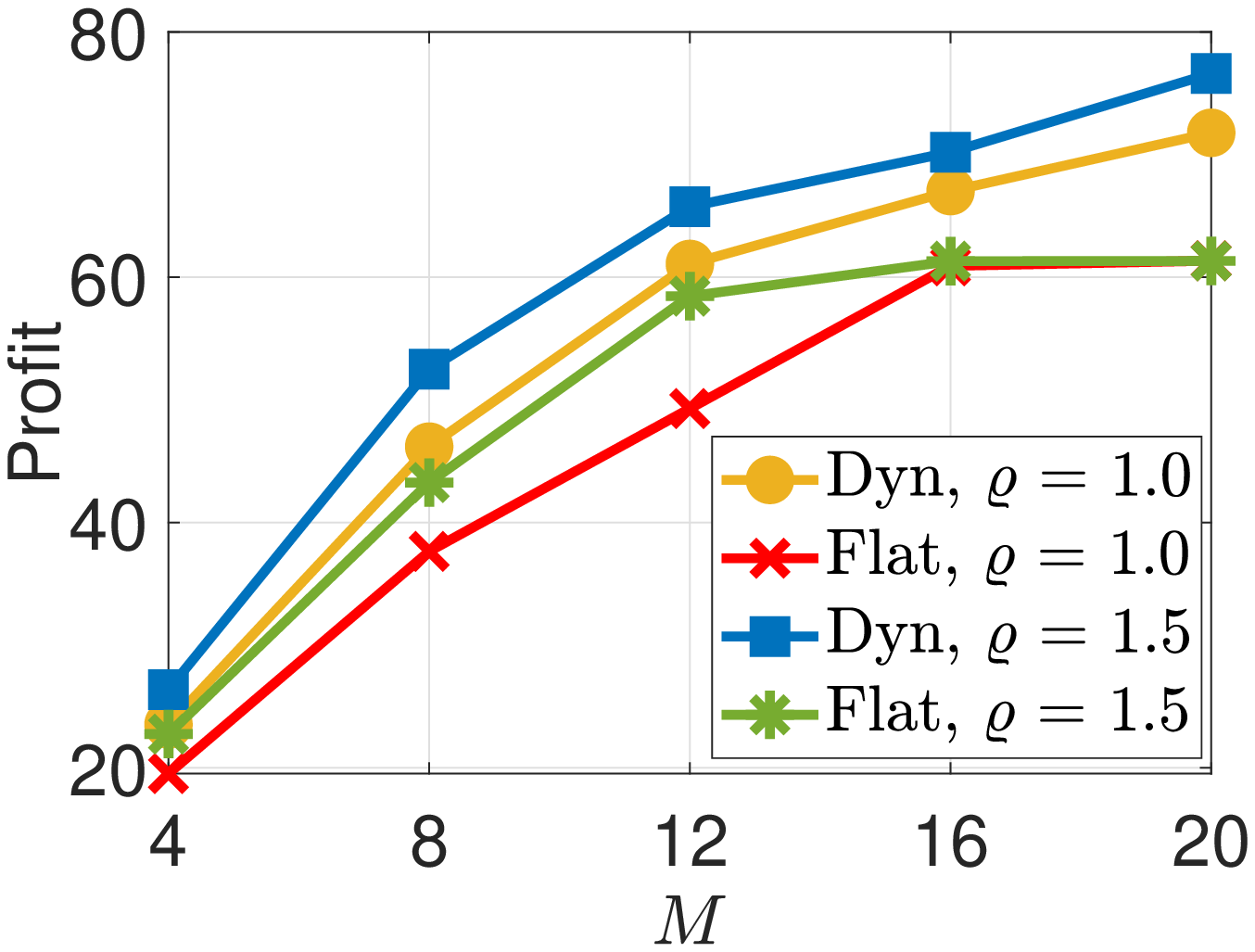}
     \label{fig:M varying cc}} 
      \vspace{-0.2cm}
      \caption{Impacts of number of APs on the system performance}
      \label{fig:Impact of M}
\end{figure}

\subsubsection{Sensitivity Analysis}
We now study the effects of various design parameters on system performance. Figs. \ref{fig:M varying R}-\ref{fig:M varying cc} 
summarizes the impact of number of APs on the system performance with varying demand, delay penalty, capacities of ENs, and cloud price by factors of $\delta$, $\Lambda$, $\gamma_0$, and $\varrho$ respectively. These figures further confirm the superior performance of \textit{Dyn}. 
Also, when there are more APs, the total workload in the system increases. Thus, we can see that the profit increases when $M$ increases due to increasing workload.

\begin{figure}[h!]
      \subfigure[Varying R and delay penalty]{\includegraphics[width=0.245\textwidth,height=0.11\textheight]{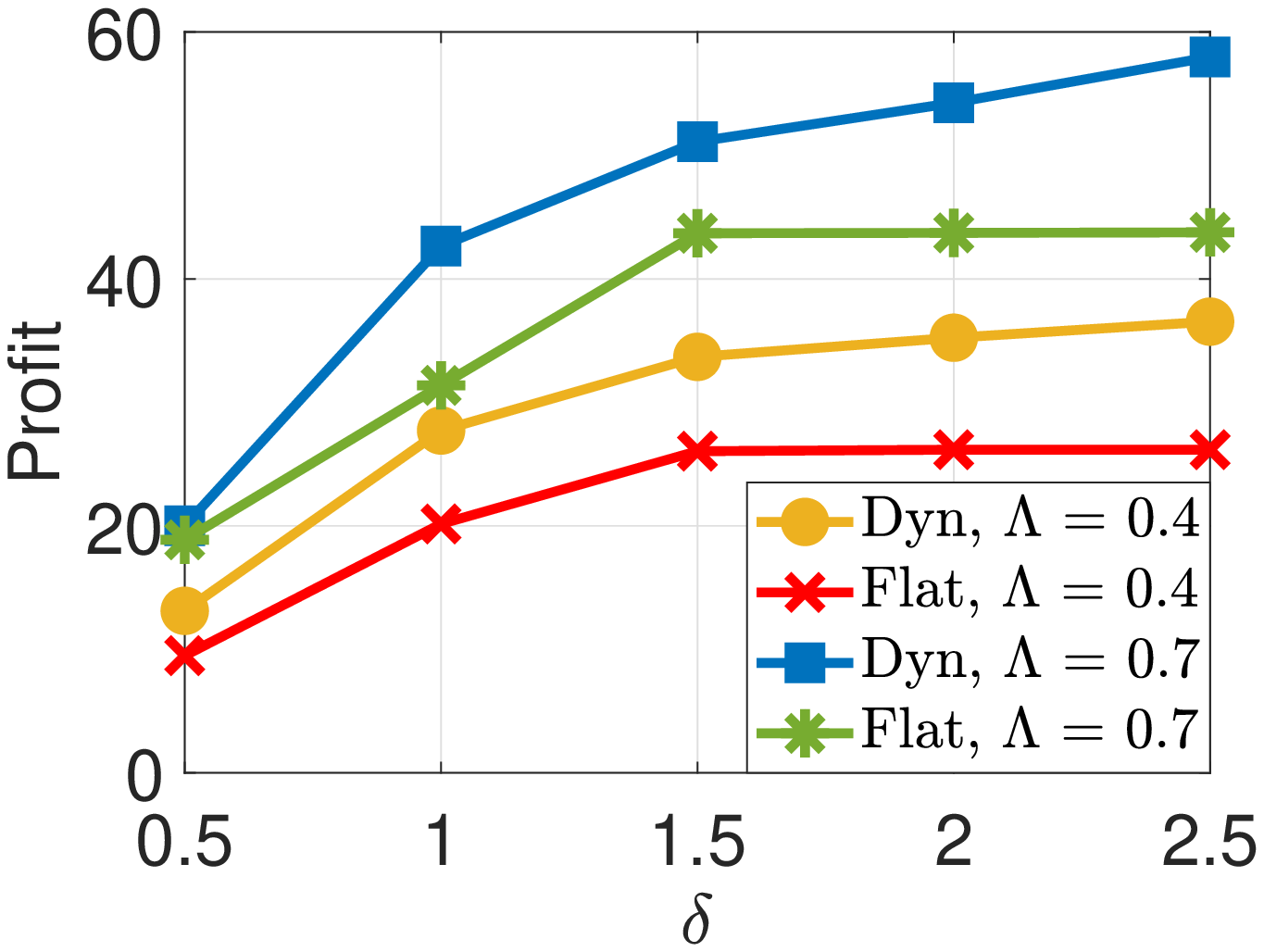}
       \label{fig:R varying w}}  \hspace*{-1.8em}
     \subfigure[Varying R and EN capacities]{\includegraphics[width=0.245\textwidth,height=0.11\textheight]{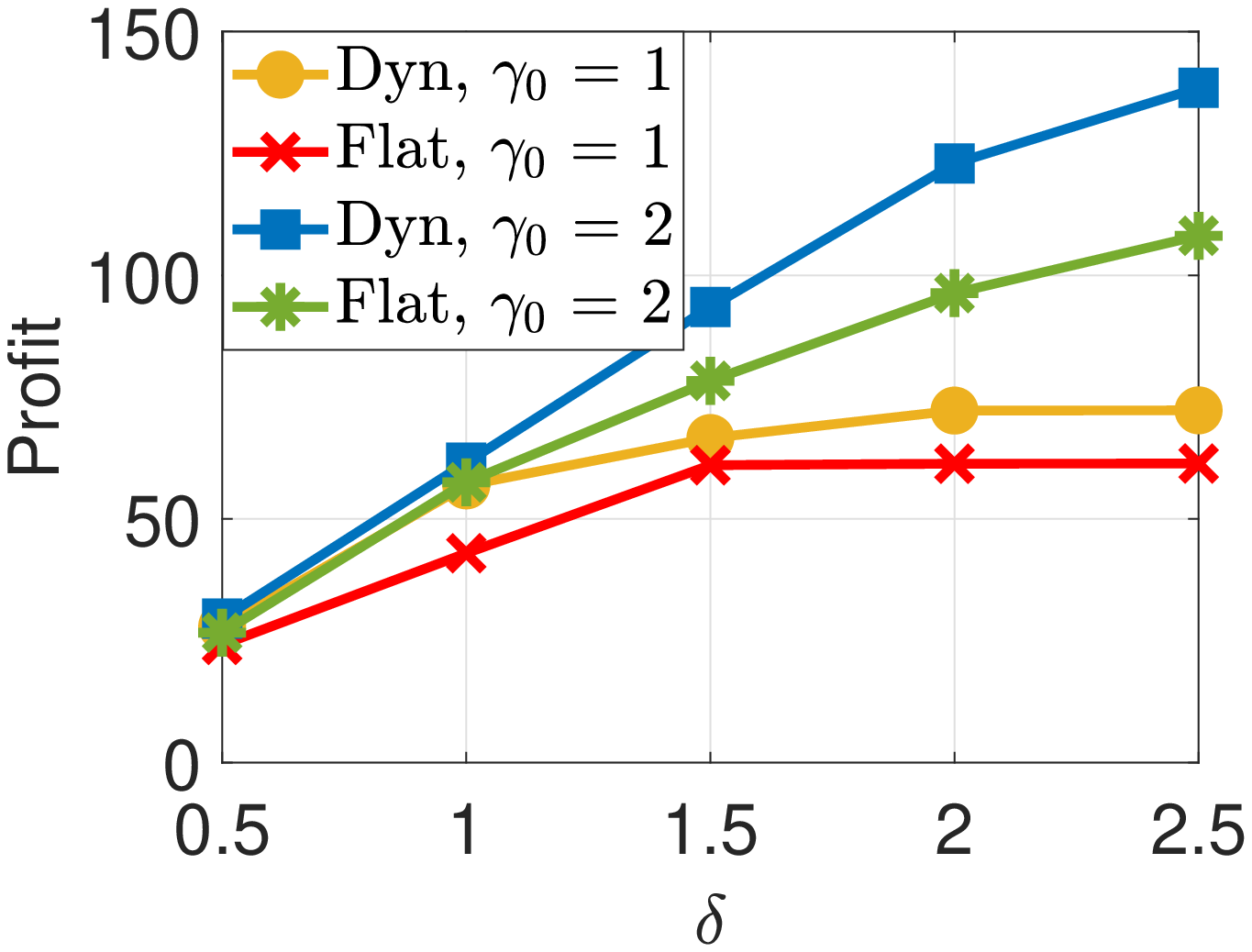}
     \label{fig:R varying C}} 
      \vspace{-0.2cm}
      \caption{Impacts of resource demand on the system performance}
      \label{fig:Impact of R}
\end{figure}

When we increase the demand, the profit further increases, as shown in Fig.~ \ref{fig:M varying w}. 
In Fig.~\ref{fig:M varying C},  the profit increases when there are more edge resources (i.e., when the EN capacities increase) because the services can buy more resources from the ENs that are beneficial for them to do offloading.  Fig.~\ref{fig:M varying cc} show that when the cloud price increases, the services will allocate more workload to the edge, which leads to increasing profits for the platform. 
Figs. \ref{fig:R varying w} and \ref{fig:R varying C} further illustrate the impacts of the resource demands of the services on the system performance. Similar to the results in Figs. \ref{fig:M varying R}-\ref{fig:M varying cc}, the profit increases as the resource demands, the delay penalty, and the  capacities of ENs increase. 
 



\begin{figure}[h!]
     \subfigure[Varying $K$ and resource demand]{\includegraphics[width=0.245\textwidth,height=0.11\textheight]{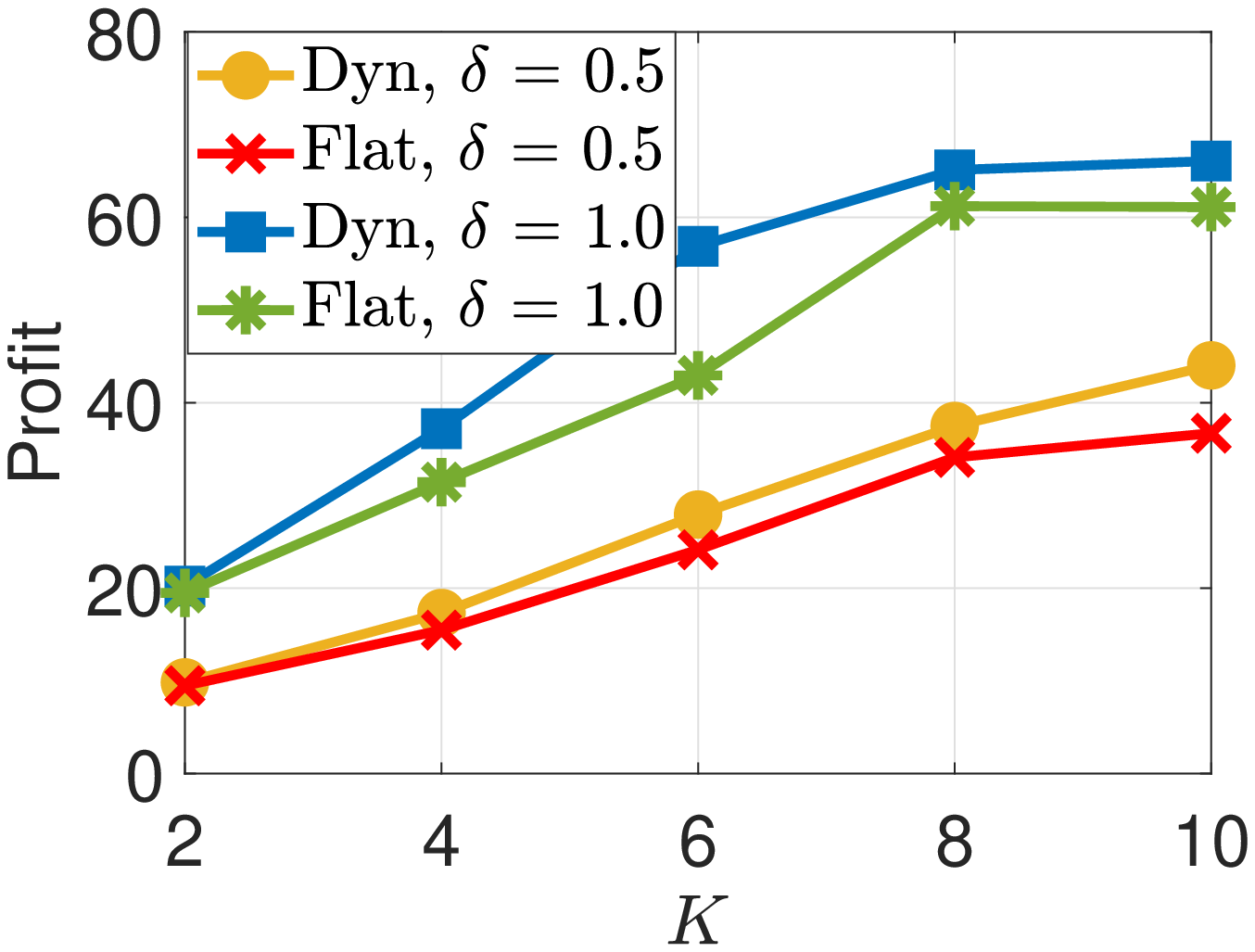} 
     \label{fig:K varying R3}} \hspace*{-1.9em}
       \subfigure[Varying $K$ and delay penalty]{\includegraphics[width=0.245\textwidth,height=0.11\textheight]{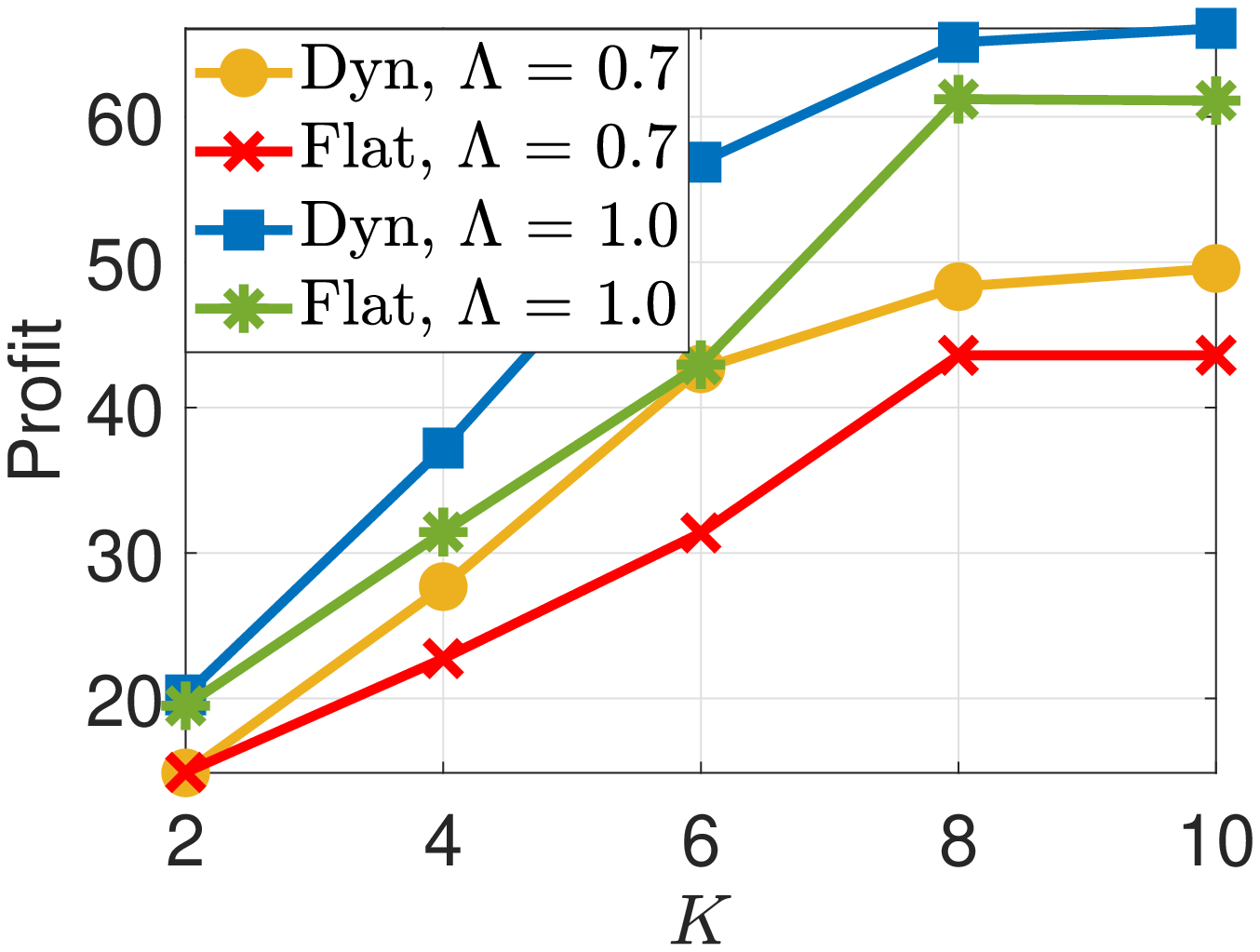} 
     \label{fig:K varying w}} 
      \subfigure[Demand scaling factor $\delta = 0.5$]{\includegraphics[width=0.245\textwidth,height=0.11\textheight]{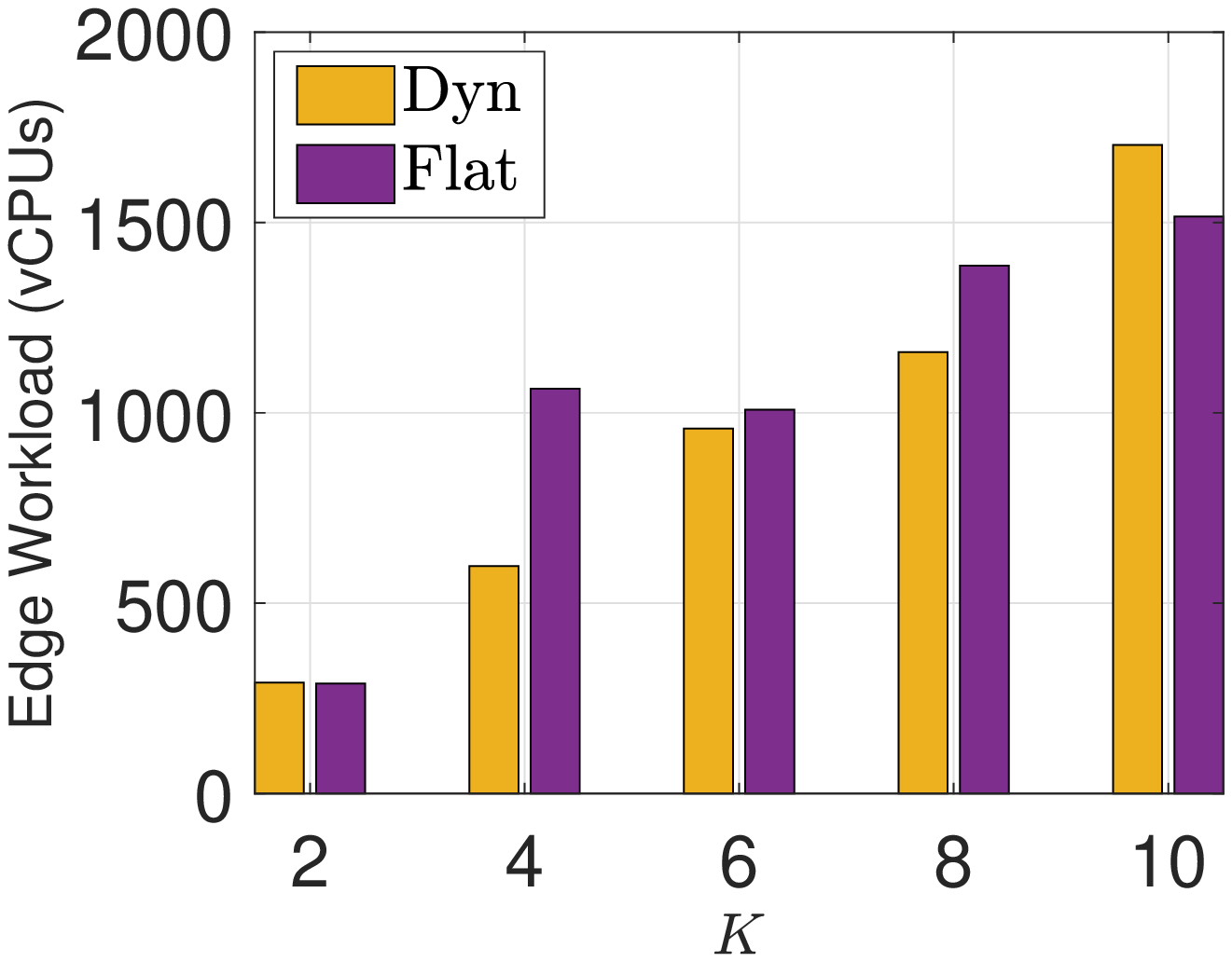}
       \label{fig:K varying R1}}  \hspace*{-2.1em}
     \subfigure[Demand scaling factor $\delta = 1$]{\includegraphics[width=0.245\textwidth,height=0.11\textheight]{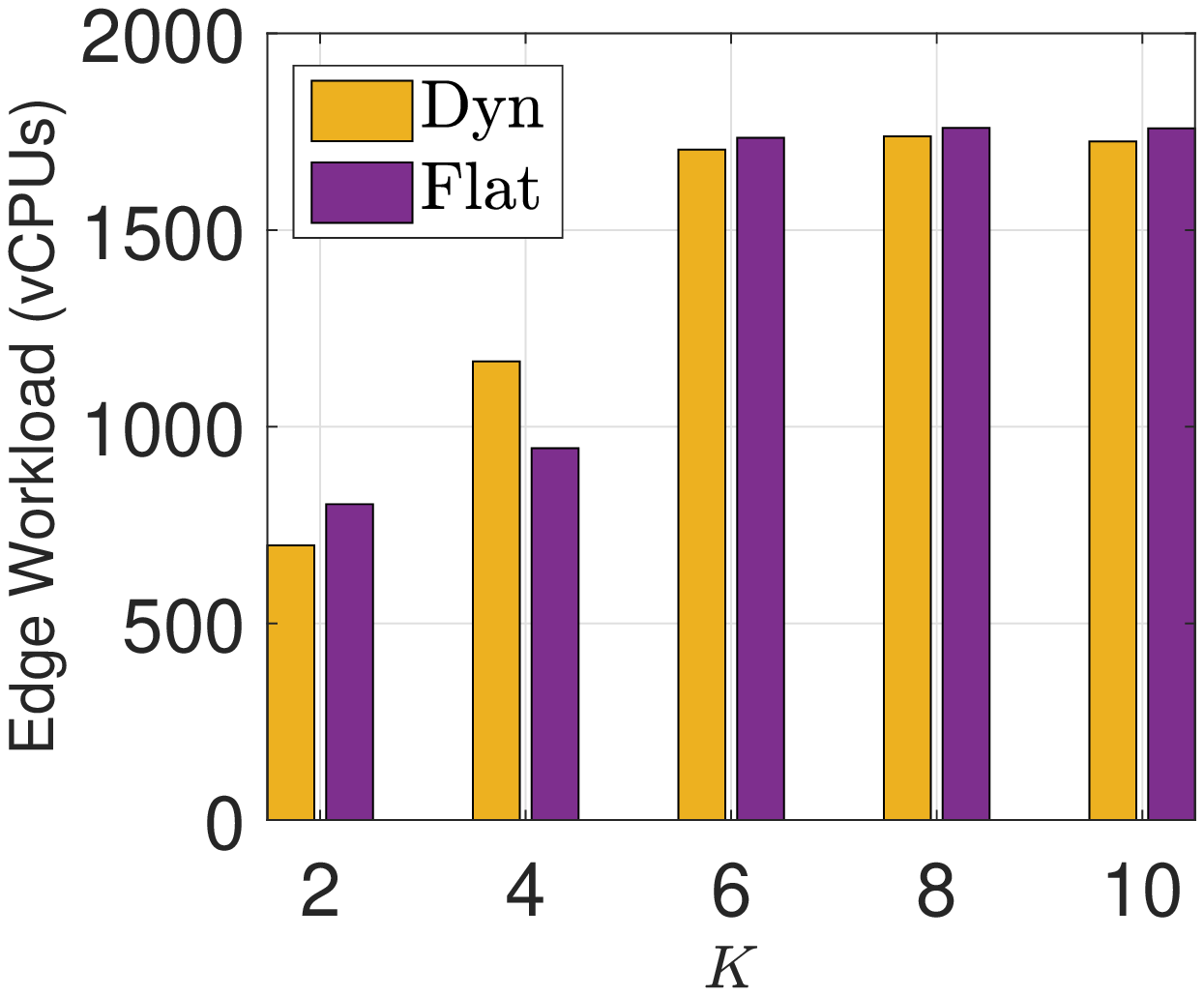} 
     \label{fig:K varying R2}} 
      \vspace{-0.2cm}
      \caption{Impacts of $K$ on the system performance}
      \label{fig:Impact of K}
\end{figure}

Figs. \ref{fig:K varying R3}-\ref{fig:K varying R2} present the impacts of the number of services on the optimal solution. 
It can be observed that the profit increases as the number of services increases. It is because when there are more services, it imposes higher resource demand in the system. Furthermore, when more services compete for edge resources, the platform can raise the edge resource prices to increase its profit. 
Figs. \ref{fig:K varying R1}-\ref{fig:K varying R2} show  the total amount of edge resource procurement of all the services. 
It is easy to see that the total workload at the ENs increases as the number of services increases due to increasing demand. When the resource demand increases (i.e., from $\delta  = 0.5$ to $\delta =1$), the amount of procured edge resources also increases. 


\section{Related Work}
\label{rel}

The emerging EC paradigm has attracted a lot of attention from the research community. 
Most of the previous work has focused on the joint optimization of communication and computational resources in mobile edge computing \cite{ymao17}.   References \cite{duongtcc} and \cite{duongton} introduce a market equilibrium approach for fair and efficient allocation of heterogeneous edge resources to  budget-constrained services. A primal-dual method for online matching between edge resources and multiple services is presented in  \cite{duongwf} to maximize the system efficiency. 
In \cite{hzha17}, Zhang \textit{et al.} combine Stackelberg game and matching theory to address the edge resource allocation problem.




Recently, the  service placement and workload scheduling problem has been studied extensively. In \cite{duongiot}, a two-stage robust optimization framework is proposed to optimize the service placement and sizing decisions for a service provider, taking into account the demand uncertainty.
Jia \textit{et al.}  \cite{mjia17}  use queuing models to jointly optimize the cloudlet placement and workload allocation decisions to minimize the system response time, considering a fixed number of cloudlets. In  \cite{lzha18}, the authors propose a ranking-based heuristic algorithm for efficient cloudlet deployment in an IoT network. 
Reference \cite{syan19} optimizes the cloudlet placement and task allocation to 
minimize the  energy consumption subject to delay constraints.  

In \cite{vfar19}, the authors present a two-timescale optimization framework to optimize service placement and request scheduling under the budget and multi-dimensional resource constraints. 
In \cite{ayou19}, A. Yousefpour \textit{et  al.}   introduce an edge service provisioning model to minimize the total system cost by  dynamically deploying and releasing applications on different ENs.
The joint service placement and request routing in mobile edge computing (MEC) is investigated in \cite{kpout19} to minimize the workload to the cloud, considering the asymmetric bandwidth requirements of the services and the limited storage capacities of ENs. 
In \cite{lwan18}, Wang \textit{et al.} examine the service  placement problem for social VR applications to minimize the total application deployment cost, including the cloudlet activation, service placement, proximity, and colocation costs.

A substantial amount of research has also been carried out on pricing design in cloud and edge networks. In \cite{hxu}, H. Xu \textit{et al.} propose a revenue maximization model and employ stochastic dynamic programming to tackle the dynamic pricing problem in an IaaS cloud. 
Similar research on joint virtual machine pricing, task scheduling, and server provisioning is studied  in \cite{jzha14} via an online profit maximization algorithm.  Reference \cite{bbae20} studies the problem of resource pricing by thoroughly analyzing several dynamic pricing schemes based on auctions and fairness-seeking properties from the perspective of game-theory and existence of unique Nash or Stackelberg equilibrium. In \cite{pcong}, the Lagrange multiplier method and a dynamic closed loop control scheme are  integrated to solve the user perceived value-based dynamic pricing problem. 

Stackelberg games and bilevel optimization have been proposed for studying resource allocation and pricing in cloud and edge computing.
In \cite{fhao19}, the authors introduce  a bilevel model and a heuristic algorithm to study the task allocation problem in a two-layer multi-community cloud/cloudlet social collaborative computational framework. The authors of \cite{mliu18} propose a Stackelberg game between a single EN and multiple mobile users, in which the former seeks to maximize revenue within capacity constraints, while the latter seeks to minimize cost performing optimal task allocation. In \cite{pqliu}, the authors present a bilevel optimization model, in which the upper-level model represents the task allocation problem and the lower-level captures the resource allocation problem, to minimize energy consumption 
under delay constraints.

Reference \cite{zxio20} studies the interaction among  cloud/edge providers that sell computing services at the upper level and a set of peer nodes called miners that decide on the service demand to be purchased at the lower level. The alternating direction method of multipliers (ADMM)  method is employed to solve this multi-leader multi-follower Stackelberg game. 
In \cite{jyan22}, the authors propose a two-stage dynamic game between wireless devices and a base station  connected to an edge server. In the first stage, the base station determines service pricing and placement decisions to maximize its profit whereas in the second stage, each device executes task offloading with the goal of reducing the service latency and cost. 

In most of the existing Stackelberg games and bilevel optimization models for cloud/edge pricing,  the follower problems are quite restrictive and have  closed-form solutions that facilitate the application of the backward induction method to find a Stackelberg equilibrium.   Some of the proposed algorithms 
are also heuristic and provide only suboptimal solutions. Unlike the previous literature, our proposed bilevel  model allows the services to be more flexible in defining their objective functions and operating constraints, which enables the follower problems to be more expressive. Furthermore, our proposed solutions are exact and give an optimal solution to the bilevel program. Our proposed model and design objective are also different from the previous work.

\section{Conclusion}
\label{conc}
In this work, we proposed a novel optimization model for joint  resource pricing, service placement, and workload allocation in EC. The interaction between an EC platform and different services was formulated as a bilevel program in which the EC platform is the leader while the services are the followers. Numerical results were presented to demonstrate the superior performance of the proposed dynamic pricing model compared to the fixed pricing schemes as well as the impacts of important system design parameters on the optimal solution.
This work opens the door to some interesting future research directions.
For example, we would like to extend the proposed model to capture various uncertainties in the system such as resource demand and component failures. We also want to study the case with multiple competing EC platforms and services. How to further improve the computational efficiency of the proposed algorithms is another interesting direction.



\bibliographystyle{IEEEtran}

\appendix

\subsection{Continuous Price}
\label{discreteAppen}
If the price $p_j$ is continuous and belongs to the range of $\big[p_j^{\sf min}, p_j^{\sf max} \big]$, we can discretize this range into  $2^{H_j}$ intervals of equal length, and the price should belong to an interval. 
The length of an interval is $ \Delta_j = \frac{ p_j^{\sf max} -  p_j^{\sf min}}{2^{H_j}}$. If $p_j$ lies in interval $l$, it can be expressed approximately  as: $p_j =  p_j^{\sf min} + \Delta_j (l - 1)$. Thus, we can express the resource price as:
\beqn
\label{pc1}
p_j =  p_j^{\sf min} + \sum_{l=1}^{ 2^{H_j} }  \Delta_j (l - 1) b_l, \\
\label{pc2}
\sum_{l=1}^{ 2^{H_j} } b_l = 1;~~b_l \in \{ 0,1\}, ~\forall l,
\eeqn
where the binary variable $b_l$ indicates if  $p_j$ lies in interval $l$. Since $p_j$ appears in several bilinear terms, we write it in the form of (\ref{pc1})-(\ref{pc2}) and further linearize the product of a binary variable and a continuous variable, as shown in (\ref{pmulinear1})-(\ref{pmulinear2}).

As $H_j$ increases, the approximation gap decreases and the accuracy increases. However when $H_j$ is large, expressing $p_j$ in forms of (\ref{pc1})-(\ref{pc2}) will create a large number of binary variables. To address this issue, we can use binary expansion to express the price more efficiently. We have:
\beqn
\label{pc3}
p_j =  p_j^{\sf min} + \sum_{l=1}^{ 2^{H_j} }  \Delta_j (l - 1), \\
\label{pc4}
l = \sum_{h = 0}^{H_j}  2^{h} b_h;~~b_h \in \{ 0,1\}, ~\forall h.
\eeqn
Therefore, by expressing $p_j$ as  (\ref{pc1})-(\ref{pc2}) or (\ref{pc3})-(\ref{pc4}), we can linearize all the bilinear terms related to $p_j$ during KKT-based reformulation and duality-based reformulation. 

\subsection{Single EN}
\label{1EN}
For the special case with only a single EN (e.g., an edge cloud serving several areas) in the system, we can solve the bilevel problem analytically. In particular, since there is only an EN, called EN1,   
the workload of any service can either be served by that EN or the cloud. This can be expressed as: 
\beqn
\label{a0}
y_0^k + y_1^k = \sum_i R_i^k,~~ \forall k.
\eeqn
By replacing $y_0^k$ by $y_1^k$ from (\ref{a0}), the budget constraint (\ref{eef6}) of service $k$ implies: 
\beqn
p_0 \big(\sum_i R_i^k - y_1^k \big) + p_1 y_1^k \leq B^k, ~~\forall k \\
\Rightarrow ~(p_1 - p_0) y_1^k \leq B^k - p_0 \sum_i R_i^k,~~ \forall k.
\label{a1}
\eeqn
There are two cases:\\
i) \textit{Case 1}: $p_1 \leq p_0$. Then, all the services will buy edge resources due to lower price and latency. If the total workload of all the services exceeds the capacity of the EN (i.e., $\sum_{i,k} R_i^k \geq C_1 $), the EN becomes overloaded and our problem becomes infeasible. On the other hand, if $\sum_{i,k} R_i^k \leq C_1 $, then workload of each service will be fully  served at the EN, i.e., $y_1^k = \sum_i R_i^k,~\forall k$. In this case, the service placement and energy costs of the platform are fixed. The amount of resource sold to the services is also fixed. Thus, the platform  should set the price as large as possible to maximize its revenue and profit, i.e., the optimal unit resource price at the EN is \beqn
p_1^{1,*} = \underset{p_1}{\max} ~\Big\{  p_1 \in \{p_1^1 , \ldots, p_1^V \}; ~p_1 \leq p_0 \Big\}.
\eeqn
The platform's profit is:
\beqn
\textit{profit}_1^* = \sum_{k,i} p_1^{1,*}  R_i^k -  \Bigg[   \big(c_1 + q_1  \frac{\sum_{k,i}  R_i^k }{C_1}\big)
+ \sum_{k}  \phi_1^k  \Bigg].
\eeqn
ii) \textit{Case 2}:  $p_1 \geq p_0$. From (\ref{a1}), we have $y_1^k \leq \alpha^k,~\forall k$, where  $\alpha^k = \frac{B^k - p_0 \sum_i R_i^k }{p_1 - p_0}$.
From the objective function of the follower problem, the cost for serving one unit of workload of service $k$ from AP $i$ at the cloud is $p_0 + w^k d_{i,0}$. Similarly, the cost for serving one unit of workload of service $k$ from AP $i$ at the EN is $p_1 + w^k d_{i,1}$. Therefore, if $p_0 + w^k d_{i,0} \leq p_1 + w^k d_{i,1}$ or equivalently $ w^k ( d_{i,0}  - d_{i,1}) \leq p_1 - p_0$ (i.e., the increased resource cost outweighs the gain from offloading), the workload of service $k$ from AP $i$ will be fully served at the cloud (i.e., $x_{i,1} = 0$ and $x_{i,0} = R_i^k$). 
On the other hand, if $p_0 + w^k d_{i,0} > p_1 + w^k d_{i,1}$, service $k$ will offload its workload at AP $i$ to the EN as much as possible.

Without loss of generality, we can index the APs such that $d_{1,1} \leq d_{2,1} \leq \ldots \leq d_{M,1}$. Then, we have $p_1 + w^k d_{M,1} \geq \ldots \geq  p_0 + w^k d_{i,0} >  p_1 + w^k d_{h,1}  \geq \ldots \geq p_1 + w^k d_{i,1}$, for some $h$. It is easy to see that the benefit of offloading increases from AP 1 to AP h. Hence, to minimize its cost, service $k$ will schedule the workload from AP 1 to AP h to the EN until the total amount of offloaded workload is equal to $\alpha^k$ due to the budget constraint. If the average delay of service $k$ is larger than $D^{k, \sf m}$, the follower problem of service $k$ is infeasible for the given value of $p_1$. 
By using the procedure above, given  $p_1$, each service can solve its corresponding follower problem analytically, without solving problem (\ref{objf})-(\ref{eef71}). 

Thus, the platform can find an optimal  price $p_1^*$ by enumerating  the set of possible prices $p_1 \geq p_0$. In particular, the platform can start by announcing the maximum price $p_1^V$. Then, each service responds by optimizing its resource procurement and workload allocation strategy using the procedure above and send $y_1^{k,*}$ to the platform. If $\sum_k y_1^{k,*} \leq C_1$, the platform computes its profit at the current price. Then, it announces the next price $p_1^{V-1}$ to the services. The procedure repeats until $\sum_k y_1^{k,*} > C_1$ at price $p_1^m$, which means the EN is over-demanded and the algorithm stops. By comparing its profits for different prices from $p_1^{m+1}$ to $p_1^V$, the platform selects the price $p_1^{2,*}$ that gives it the highest profit, called $\textit{profit}_2^*$.

Finally, if  case 1 and case 2 are both feasible, the platform will choose the higher optimal profit. Specifically, if $\textit{profit}_1^* < \textit{profit}_2^*$, the optimal price is $p_1^* = p_1^{2,*}$. Otherwise,   $p_1^* = p_1^{1,*}$.  

\end{document}